\documentclass[a4paper,fleqn,usenatbib]{mnras}

\usepackage{newtxtext,newtxmath}

\usepackage[T1]{fontenc}
\usepackage{ae,aecompl}

\usepackage{graphicx}	
	\graphicspath{{figures/}}
	
\usepackage{amsmath}	
\usepackage{amssymb}	

\usepackage{mathtools}

\DeclareMathOperator{\sign}{sign}

\usepackage{hyperref} 
	\hypersetup{colorlinks,citecolor=blue,linkcolor=blue,urlcolor=blue}
	\usepackage{etoolbox}
	
\usepackage[capitalise]{cleveref}
	\crefname{equation}{Equation}{Equations}
	\crefname{figure}{Figure}{Figures}
	\crefname{table}{Table}{Tables}
	
	\newcommand{\crefalt}[1]{\namecref{#1}~\ref{#1}}
	
\usepackage{ulem}
	
\newcommand{\R}{R}							
\newcommand{\Rg}{\R_{\mathcal{G}}}				
\newcommand{\zIF}{z_{\mathrm{IF}}}					

\newcommand{\rhog}{\rho_{\mathrm{g}}}				
\newcommand{\rhogO}{\rho_{\mathrm{g,0}}}			
\newcommand{\rhogn}{\rho_{\mathrm{g,n}}}			
\newcommand{\rhogi}{\rho_{\mathrm{g,i}}}				
\newcommand{\rhod}{\rho_{\mathrm{d}}}				
\newcommand{\rhograin}{\rho_{\mathrm{grain}}}		
\newcommand{\rhoeff}{\rho_{\mathrm{eff}}}			

\newcommand{\Vg}{u}							
\newcommand{\vg}{\Vg}							
\newcommand{\vgO}{\Vg_0}						
\newcommand{\vgn}{\Vg_{\mathrm{n}}}				
\newcommand{\vgi}{\Vg_{\mathrm{i}}}				
\newcommand{\vK}{v_{\mathrm{K}}}					
\newcommand{\Mach}{\mathcal{M}}					
\newcommand{\MachK}{\Mach_{\mathrm{K}}}			
\newcommand{\MachO}{\Mach_0}					
\newcommand{\Machn}{\Mach_{\mathrm{n}}}			

\newcommand{\Vd}{v}						    	
\newcommand{\vdi}{\Vd_{\mathrm{i}}}				
\newcommand{\vd}{\Vd}							
\newcommand{\vgpar}{\Vg_\parallel}              
\newcommand{\vgnpar}{\Vg_{\mathrm{n}\parallel}} 
\newcommand{\vgipar}{\Vg_{\mathrm{i}\parallel}} 
\newcommand{\vgiperp}{\Vg_{\mathrm{i}\perp}}    
\newcommand{\vgnperp}{\Vg_{\mathrm{n}\perp}}    
\newcommand{\vgperp}{\Vg_\perp}                 

\newcommand{\ts}{t_{\mathrm{s}}}				
\newcommand{\tsn}{t_{\mathrm{s,n}}}				
\newcommand{\tsi}{t_{\mathrm{s,i}}}				
\newcommand{\St}{\mathrm{St}}					
\newcommand{\Sti}{\St_{\mathrm{i}}}				
\newcommand{\StIF}{\St(\zIF)}			        
\newcommand{\Stcrit}{\St_{\mathrm{crit}}}		
\newcommand{\Stmax}{\St_{\mathrm{max}}}	    	
\newcommand{\Stimax}{\St_{\mathrm{i,max}}}		
\newcommand{\Stlimit}{\St_{\mathrm{limit}}}	    

\newcommand{\smaxmrn}{s_{\mathrm{MRN}}}		
\newcommand{\eps}{\varepsilon}					
\newcommand{\scrit}{s_{\mathrm{crit}}}          
\newcommand{\smax}{s_{\mathrm{max}}}            
\newcommand{\slimit}{s_{\mathrm{limit}}}        

\newcommand{\Dd}{\mathcal{D}_{\mathrm{d}}}			
\newcommand{\Sc}{\mathrm{Sc}}					

\newcommand{\OmegaK}{\Omega_{\mathrm{K}}}		
\newcommand{\OmegaKau}{\OmegaK{}_{,1\!\au}}		

\newcommand{\Sigmag}{\Sigma_{\mathrm{g}}}			
\newcommand{\Sigmagau}{\Sigmag{}_{,1\!\au}}			

\newcommand{\cs}{c_{\mathrm{s}}}					
\newcommand{\csn}{c_{\mathrm{s,n}}}				
\newcommand{\csi}{c_{\mathrm{s,i}}}				
\newcommand{\csau}{\cs{}_{,1\!\au}}					

\newcommand{\Hg}{H}							
\newcommand{\Hgau}{\Hg{}_{1\!\au}}				

\newcommand{\Phii}{\Phi_{\mathrm{i}}}				
\newcommand{\iIF}{i_{\rm IF}}                   

\newcommand{\Mstar}{M}							

\newcommand{\G}{\mathcal{G}}					
\newcommand{\F}{\mathcal{F}}					
\newcommand{\Fadv}{\F_{\rm adv}}				
\newcommand{\Fdiff}{\F_{\rm diff}}				
\newcommand{\f}{f}						        
\newcommand{\mH}{m_{\mathrm{H}}}				

\newcommand{\gram}{\; {\mathrm{g}}}				
\newcommand{\cm}{\; {\mathrm{cm}}}				
\newcommand{\mum}{\; \mu {\mathrm{m}}}			
\newcommand{\nm}{\; {\mathrm{nm}}}				
\newcommand{\pers}{\; {\mathrm{s}}^{-1}}				
\newcommand{\kms}{\; {\mathrm{km}} \pers}			
\newcommand{\au}{\; {\mathrm{au}}}					
\newcommand{\Myr}{\; {\mathrm{Myr}}}				
\newcommand{\ionflux}{\; {\mathrm{photons}} \pers}		

\newcommand{\unsim}{\mathord{\sim}}
\newcommand{\ungtrsim}{\mathord{\gtrsim}}

\defcitealias{Owen/Ercolano/Clarke/2011a}{OEC11}
\defcitealias{Hutchison/etal/2016}{HPLM16a}
\defcitealias{Hutchison/Laibe/Maddison/2016}{HLM16b}
\defcitealias{Clarke/Alexander/2016}{CA16}

\setcitestyle{notesep={; }}

\title[Dust delivery and entrainment]{Dust delivery and entrainment in photoevaporative winds}

\author[Hutchison \& Clarke]{
Mark A. Hutchison,$^{1,2,3}$\thanks{E-mail: markahutch@gmail.com}
Cathie J. Clarke,$^{4}$
\\
$^{1}$Universit{\"a}ts-Sternwarte, Ludwig-Maximilians-Universit{\"a}t  M{\"u}nchen, Scheinerstr. 1, 81679 M{\"u}nchen, Germany \\
$^{2}$Physikalisches Institut, Universit{\"a}t Bern, Gesellschaftstrasse 6, 3012 Bern, Switzerland \\
$^{3}$Institute for Computational Science, University of Z{\"u}rich, Winterthurerstrasse 190, CH-8057 Z{\"u}rich, Switzerland \\
$^{4}$Institute of Astronomy, Madingley Road, Cambridge CB3 OHA, UK
}

\date{Accepted XXX. Received YYY; in original form ZZZ}

\pubyear{2020}

\begin{document}
\label{firstpage}
\pagerange{\pageref{firstpage}--\pageref{lastpage}}
\maketitle

\begin{abstract}
    We model the gas and dust dynamics in a turbulent protoplanetary disc undergoing extreme-UV photoevaporation in order to better characterise the dust properties in thermal winds (e.g. size distribution, flux rate, trajectories).
    Our semi-analytic approach allows us to rapidly calculate these dust properties without resorting to expensive hydrodynamic simulations. 
    We find that photoevaporation creates a vertical gas flow within the disc that assists turbulence in supplying dust to the ionisation front. We examine both the {\it delivery} of dust to the ionisation front and its subsequent {\it entrainment} in the overlying wind.  We derive a simple analytic criterion for the maximum grain size that can be entrained and show that this is in good agreement with  the results  of previous  simulations where photoevaporation is driven by a range of radiation types.  We show that, in contrast to the case for magnetically  driven winds, we do not expect large scale dust transport within the disc to be effected by photoevaporation. We also show that the maximum size of  grains that can be entrained in the wind ($\smax$) is around an order of magnitude larger than the maximum size of grains that can be delivered to the front by advection alone ($\scrit  \lesssim 1 \mum$ for Herbig Ae/Be stars and $\lesssim 0.01 \mum$ for T Tauri stars).
    We further investigate how larger grains, up to a limiting size $\slimit$, can be delivered to the front by turbulent diffusion alone. In all cases, we find $\smax \ungtrsim \slimit$ so that we expect that any dust that is delivered to the front can be entrained in the wind and that most entrained dust follows trajectories close to that of the gas.
\end{abstract}




\section{Introduction}
\label{sec:introduction}

Planets are born in the nurturing cocoon of gas and dust created by single, or potentially binary \citep{Quintana/etal/2007}, parent stars early in their own development. While the circumstances surrounding the birth of each planet may be unique -- from fragmentation via gravitational instability \citep[e.g.][]{Kuiper/1951,Boss/1997,Inutsuka/Machida/Matsumoto/2010,Nayakshin/2010} to streaming instability of large dust reservoirs \citep[e.g.][]{Youdin/Goodman/2005,Youdin/Johansen/2007,Johansen/Youdin/2007} followed by core accretion \citep[e.g.][]{Safronov/1969,Pollack/etal/1996,Ikoma/Nakazawa/Emori/2000}  -- all planets, regardless of how they were born, inevitably experience the death of the protoplanetary disc phase early in their evolution \citep[$\lesssim 5$--$10 \Myr$; e.g.][]{Haisch/Lada/Lada/2001,Hernandez/etal/2007,Mamajek/2009}. These discs are crucial to the early development of planets because they provide the fodder for growth and a driver of radial migration \citep[e.g.][]{Goldreich/Tremaine/1980,Ward/1986,Ward/1997,Tanaka/Takeuchi/Ward/2002}. Therefore, putting constraints on the dispersal mechanisms and/or dispersal rates of gas and dust in these discs is important for understanding the architectures of planetary systems we observe today \citep[e.g][]{Alexander/Pascucci/2012,Mordasini/etal/2012}.

Many dispersal mechanisms are likely to play a role in the evolution of protoplanetary discs: viscous accretion \citep{Shakura/Sunyaev/1973,Lynden-Bell/Pringle/1974}, planet-disc interactions \citep{Calvet/etal/2005}, grain growth \citep{Dullemond/Dominik/2005}, photophoresis \citep{Krauss/etal/2007}, MRI-driven winds \citep{Suzuki/Inutsuka/2009}, binary star interactions \citep{Marsh/Mahoney/1992}, MRI-driven dust depleted flows \citep{Chiang/Murray-Clay/2007}, magneto-thermal winds \citep{Bai/etal/2016},  and photoevaporation \citep{Clarke/Gendrin/Sotomayor/2001}. It is however often assumed that photoevaporation is responsible for the final clearing of the disc, in part because of its success in explaining the {\it rapid} transition between disc-bearing and disc-less stars \citep{Owen/Ercolano/Clarke/2011b,Ercolano/etal/2015}.
Predictions of gas mass-loss rates \citep[e.g.][]{Ercolano/etal/2008,Gorti/Dullemond/Hollenbach/2009,Owen/Clarke/Ercolano/2012,Picogna/etal/2019} and their effect on disc evolution \citep{Ercolano/Rosotti/2015,Ercolano/Pascucci/2017,Carrera/etal/2017,Jennings/Ercolano/Rosotti/2018} have been studied in great detail.

While dust was initially  assumed to be blown out with the gas \citep{Johnstone/Hollenbach/Bally/1998,Matsuyama/Johnstone/Hartmann/2003,Adams/etal/2004}, \citet {Throop/Bally/2005} provided the first study of the differential removal of dust and gas in a photoevaporative wind. 
With the majority of dust mass in discs locked up in large grains near the disc mid-plane and traced by its sub-mm thermal emission, the best prospect of using dust to trace photoevaporative winds is through scattered light imaging above the disc photosphere.
\citet{Owen/Ercolano/Clarke/2011a} found that dust pulled from the disc can leave a characteristic imprint on the surrounding environment that can be detected in older edge-on discs.
They further found that the morphology of their scattered light images depended on the maximum grain size pulled into the wind at different radii. 
\citet{Takeuchi/Clarke/Lin/2005} provided the first order-of-magnitude estimate for the maximum grain size in winds. Later
\citet[][hereafter, \citetalias{Hutchison/Laibe/Maddison/2016}]{Hutchison/Laibe/Maddison/2016}  found an analytic relation for the maximum entrainable grain size in vertical winds and postulated that the equation could be inverted to give a lower limit on the gas mass-loss rate if observations could constrain the grain sizes that were being pulled from the disc. While these examples are illustrative of how dust may be used to constrain disc winds,
each of these studies suffers from some limitation in modelling the delivery and/or subsequent entrainment of solids in the wind.

To date, there have been five two-phase (gas+dust) photoevaporation wind models presented in the literature -- none of which provide the robust understanding that we need to confidently set constraints on dispersal mechanisms using dust measurements:
\begin{enumerate}
	\item	\citet{Owen/Ercolano/Clarke/2011a} used a full grain-size distribution for the dust \citep{Mathis/Rumpl/Nordsieck/1977}, a self-consistent ionisation front prescription for the gas \citep{Hollenbach/etal/1994}, and checked for dust entrainment along gas streamlines in a 2D extreme ultraviolet (EUV) driven wind. The disadvantage with their model is that the flow originates from an infinitely-thin, perfectly-mixed disc and the dust in the wind was required to have the same streamlines as the gas.
	\item	\citet{Facchini/Clarke/Bisbas/2016} also used a realistic grain-size distribution and, with the aid of the {\sc 3D-PDR} code \citep{Bisbas/etal/2012}, self-consistently solved the hydrodynamics of gas and dust in a radial flow from the outer disc. In addition to only being 1D, their model is restricted to external photoevaporation caused by far-UV (FUV) sources.
	\item	\citet[][hereafter, \citetalias{Hutchison/etal/2016}]{Hutchison/etal/2016}
	used smoothed particle hydrodynamics to self-consistently model the two-phase hydrodynamics with a single grain size in thin vertical atmospheres undergoing EUV photoevaporation. However, their 1D model neglects rotational effects in the wind, the ionisation front location is a free parameter, and the disc is laminar (i.e. settling dust never reaches a steady state). 
	\item	\citetalias{Hutchison/Laibe/Maddison/2016} developed a fast semi-analytic model based on \citetalias{Hutchison/etal/2016} that included turbulent mixing within the disc but treated the gas as being static below the ionisation front with no mechanism for advecting the dust upwards. Their model also suffers from having a 1D flow and lacks a self-consistently set ionisation front.
	\item \citet{Franz/etal/2020} used a 2D hydrodynamical disc model irradiated by X-ray and EUV spectra (XEUV) from a central T Tauri star to measure entrainment in winds using tracer particles of various sizes inserted at the disc surface. 
	Since this study only tracks the trajectories of dust particles within the wind, it cannot assess dust mass-loss rates, since the delivery
    of dust into the wind, and how this depends on the grain size and the properties of the underlying turbulent disc, is not addressed.
\end{enumerate}

One of the common themes emerging from the work of \citetalias{Hutchison/etal/2016} and \citetalias{Hutchison/Laibe/Maddison/2016} is that the local conditions in the disc beneath the flow can limit the dust distribution in the wind, despite the wind being able to entrain larger grains. Therefore, accurate diagnostics of dust in winds (such as size-distribution, mass-loss, and density) require models that can self-consistently link the gas and dust properties back to their respective reservoirs in the disc. Two-phase, 3D radiation hydrodynamic simulations can potentially provide the best constraints on dusty winds, but as such models are not readily available, we focus on combining and improving upon the following elements from previous studies:
\begin{enumerate}
    \item We set the ionisation front location as a function of incident EUV flux \citep{Hollenbach/etal/1994}.
    \item Using an adaptation of the self-similar 2D wind solution of \citet{Clarke/Alexander/2016} that accounts for the finite height of the disc surface, we set the gas velocity above the ionisation front so as to enable the 2D ionised wind solution to pass smoothly through its critical point.
    \item Below the ionisation front, we solve for the vertical profile of dust species of various sizes including gravitational settling, turbulent diffusion, and hydrodynamic drag (from the vertical gas flow feeding the base of the wind).
    \item We approximate and spatially resolve the physical width of the ionisation front to provide a smooth transition between the turbulent disc and the laminar wind.
\end{enumerate}
The combination of the above features allows us to track the flow of dust from the disc mid-plane into the wind, thereby coupling the dusty wind to the disc. Our goal is to investigate the size range, flux, and subsequent trajectory of dust passing through the ionisation front. In what follows we will refer to processes by which grains arrive at the ionisation front as {\it delivery} and the subsequent dynamical evolution of the dust in the ionised wind flow as {\it entrainment}.

The structure of the paper is as follows. In \cref{sec:methods} we describe the vertical structure of the gas disc and how we solve for the steady state dust distribution  and corresponding dust flux, as a function of grain size.
In \cref{sec:results} we describe results relating to both dust delivery and entrainment. \cref{sec:implications} discusses the implications of our results for dusty photoevaporating discs and \cref{sec:conclusions} summarises our conclusions.

\section{Methods}
\label{sec:methods}

We proceed by building a gas model for the disc and wind. Then we add dust to the disc mid-plane and solve the fluid equations for the dust assuming the backreaction on the gas is negligible \citepalias[for justification, see][]{Hutchison/etal/2016}. In what follows, we use the subscripts $_\mathrm{g}$ and  $_\mathrm{d}$ to distinguish between gas and dust variables with the same name.

We consider a disc whose internal structure is approximately described by the following power-law parameterisations \citep[see, e.g.,][]{Laibe/Gonzalez/Maddison/2012}:
\begin{align}
	\rhogO & =  \frac{\Sigmag}{\sqrt{2\pi} \Hg},
	\label{eq:disc_rhogmid}
\\
	\Sigmag & \propto \left( \frac{\R}{\Rg} \right)^{-p},
	\label{eq:disc_sigmag}
\\
	\Hg & \propto \left( \frac{\R}{\Rg} \right)^{3/2- q/2},
	\label{eq:disc_hg}
\\
	\cs &  = \OmegaK \Hg \propto \left( \frac{\R}{\Rg} \right)^{-q/2},
	\label{eq:disc_cs}
\end{align}
where $\rhogO$ is the mid-plane density, $\Sigmag$ is the local surface density, $\OmegaK = (\G \Mstar/\R^3)^{1/2}$ is the Keplerian orbital frequency and $p$ and $q$ are power-law exponents respectively controlling the surface density and temperature profiles of the disc. In this paper we assume the following generic reference values for our disc: $p = 1$, $q = 0.5$, $\Mstar = M_\odot$, $\Sigmagau = 100 \gram \cm^{-3}$, $ \Hgau = 0.05 \au$, and $\csau = \OmegaKau \, \Hgau \approx 1.5 \kms$.
Later, however, we will modify the 
radial power-law profiles for the sound speed and disc scale height in order to be consistent with the height of the ionisation front and the vertical flow it sets up within the disc.\footnote{Also note that by including the flow in the disc, the mid-plane density, surface density, and the disc scale-height all deviate very slightly from their respective hydrostatic relations. However, given that the changes to the disc structure occur almost exclusively in the low-density regions near the disc surface, these deviations are negligible.}

\subsection{Gas flow}
\label{sec:gas_flow}
We begin by identifying the physical constraints at the ionisation front that are needed to connect the gas flow in the disc to the wind.

\subsubsection{ Jump conditions at the ionisation front}
\label{sec:jump_conditions}
Conservation of mass and momentum across the ionisation front allows us to relate the density ($\rhog$) and the perpendicular velocity ($\vgperp$) in the neutral disc with corresponding quantities in the ionised wind:
\begin{align}
	\rhogn \vgnperp & = \rhogi \vgiperp,
	\label{eq:RH_mass_conserv}
\\
	\rhogn \csn^2 + \rhogn \vgnperp^2 & = \rhogi \csi^2 + \rhogi \vgiperp^2,
	\label{eq:RH_mom_conserv}
\end{align}
where $\cs$ is the sound speed of the gas and the subscripts $_{\mathrm{n}}$ and $_{\mathrm{i}}$ refer to the neutral and ionised sides, respectively. Setting the sound speed on the neutral side to be $\csn=\cs$ (which is assumed to be a function of radius only), we can solve \cref{eq:RH_mass_conserv,eq:RH_mom_conserv}
for $\rhogn$ and $\vgnperp$. After some manipulation we find
\begin{align}
	\vgnperp & = \dfrac{\csi^2 + \vgiperp^2  \pm \sqrt{\left( \csi^2 + \vgiperp^2 \right)^2 - 4 \cs^2 \vgiperp^2}} {2 \vgiperp},
	\label{eq:vgn}
\\
	\rhogn  & = \dfrac{\rhogi \vgiperp}{\vgnperp} = \frac{\rhogi \left[ \csi^2 + \vgiperp^2  \mp \sqrt{\left( \csi^2 + \vgiperp^2 \right)^2 - 4 \cs^2 \vgiperp^2}  \right] } {2 \cs^2},
	\label{eq:rhogn}
\end{align}
where we take the negative sign in \cref{eq:vgn} and the positive sign in \cref{eq:rhogn} to be the physical solutions because the velocity in the disc must be smaller than the wind speed.

If the gas flow is perpendicular to the ionisation front (i.e. if $\vgi = \vgiperp$ and $\vgn = \vgnperp$) then \cref{eq:vgn,eq:rhogn} are sufficient to constrain the physical properties of our flow. 
However, when the ionisation front is tilted by angle $\iIF$  with respect to the flow, as assumed in this study, there is an additional constraint on the gas velocity; namely, that the velocity parallel to the front remains constant across the discontinuity: $\vgipar = \vgnpar = \vgpar$. Assuming that we know the wind density and total velocity at the base of the ionised flow as well as the sound speed in the disc and wind, that leaves us with five unknown variables. The neutral quantities $\vgnperp$ and $\rhogn$ continue to depend on $\vgiperp$ according to \cref{eq:vgn,eq:rhogn} while the remaining unknown velocities ($\vgiperp$, $\vgn$, and $\vgpar$) can all be related using trigonometric relations, thereby closing the system of equations.

With some effort, it can be shown that $\vgpar$ satisfies the following polynomial equation
\begin{equation}
    c_6 \vgpar^6 + c_4 \vgpar^4 + c_2 \vgpar^2 + c_0 = 0,
    \label{eq:vgpar}
\end{equation}
with coefficients
\begin{align}
    c_6 & = \cot^2{\iIF} \csc^2{\iIF},
    \label{eq:c6}
\\
    c_4 & =  \left[ 2 \left( \cs^2 - \csi^2 \right) - \vgi^2 \left(1 + \csc^2{\iIF} \right) \right] \cot^2{\iIF},
    \label{eq:c4}
\\
    c_2 & = \cs^4 + \left[ \left( \csi^2 + \vgi^2 \right)^2 - 2 \cs^2 \vgi^2 \right] \cot^2{\iIF},
    \label{eq:c2}
\\
    c_0 & =  - \cs^4 \vgi^2.
    \label{eq:c0}
\end{align}
Although the roots of \cref{eq:vgpar} can be expressed analytically in closed form, they are too extensive to be of practical use. Instead we solve for $\vgpar$ numerically, after which $\vgiperp$ and the remaining quantities are easily calculated. For a full derivation of \cref{eq:vgn,eq:rhogn,eq:vgpar,eq:c6,eq:c4,eq:c2,eq:c0}, see \cref{sec:jump_cond_derivation}.

\subsubsection{Determination of the velocity at the base of the ionised flow}
\label{sec:ui}

In order to determine $\vgi$ we follow the approach of  \citet{Clarke/Alexander/2016} who computed the structure of thermally driven axisymmetric winds under the assumption of self-similarity.
Clarke \& Alexander demonstrated that there is a maximum launch velocity for which the flow structure contains a critical point at the sonic point and that this maximum velocity agrees well with the launch velocity found in hydrodynamical solutions of disc winds. While Clarke \& Alexander only treated the case that the wind is launched from the disc mid-plane, we here consider the more realistic scenario of a wind launched from an inclined surface. In order to preserve self-similarity in the wind, we assume this surface makes a constant angle $\iIF = 0.3~{\rm rad}$ to the horizontal, as detailed in \cref{sec:IF_location}. In this study, we keep $\iIF$ fixed for all parameter combinations in order to make comparisons between simulations easier.

We find the launch velocity by sampling $\vgi \in [0,\,\cs]$ and calculating the parallel and perpendicular components $\vgipar$ and $\vgiperp$ according to \cref{sec:jump_conditions}. These components determine the initial launch angle of the flow. Using these initial conditions, we then integrate the equations describing the topology, density, and velocity structure of a self-similar flow.\footnote{In connecting the disc to the wind, self-similarity is not strictly maintained. Because the disc is not globally isothermal, the radial dependence in \cref{eq:disc_cs} weakly influences the launching angle of the wind. Fortunately, the large temperature jump at the ionisation front ensures that $\vgipar \ll \vgiperp$ and that the ionised wind emerges {\it nearly} perpendicular to the ionisation front. The departure from self-similarity is small (variations of $<1\%$) and can safely be ignored.} Finally, we iterate on this procedure until we find the maximum velocity $\vgi$ that permits a transonic flow that extends to infinity.

\subsubsection{Ionised boundary}
\label{sec:ionised_boundary}

The ionised density at the base of the wind is independently determined by solving the ionisation-recombination balance for an isothermal disc irradiated by EUV radiation. Using the weak stellar wind model from \citet{Hollenbach/etal/1994}, the density of the ionised wind in the inner few au of the disc can be expressed analytically by
\begin{align}
	\rhogi & = \mH C_0 \left( \dfrac{3 \Phii} {4 \pi \alpha_2 \R^3} \right)^{1/2},
	\label{eq:hollenbach_density}
\end{align}
where $\R$ is the cylindrical disc radius measured from a central star of mass $\Mstar$, $\mH$ is the mass of hydrogen,
           $\alpha_2 = 2.6 \times 10^{-13} \cm^3 \pers$ is the recombination coefficient for all states except to the ground state (case B),
           $\Phii$ is the stellar EUV luminosity, and
           $C_0 = 0.2$ is the numerical factor used by \citet{Hollenbach/etal/1994} to bring their analytical and numerical results into agreement.
In this paper, we will only consider two extremes for the EUV luminosity, $\Phii = [10^{41},\,10^{45}] \ionflux$ (hereafter referred to as $\Phi_{41}$ and $\Phi_{45}$), and assume that the resulting wind has an isothermal temperature of $T = 10~000\,{\rm K}$ or, equivalently, a sound speed $\csi \sim 10 \kms$. Note $\Phi_{41}$ corresponds to a typical T Tauri star whereas $\Phi_{45}$ would represent an early type Herbig Be star \citep{Gorti/Dullemond/Hollenbach/2009} with a larger mass. Because changing the stellar mass clouds the interpretation of the results, we choose to keep $M$ constant as we vary $\Phii$.

Although \cref{eq:hollenbach_density} formally only applies to disc radii between where the disc scale height equals the radius of the star and the so-called gravitational radius $\Rg = \G \Mstar / \cs^2$, we have adopted it as a general relation for our disc. We caution that the resulting mass-loss rate diverges to large radii as $\sqrt{\R}$ and that, in real systems, the radial range over which this approximation holds is limited by the finite energy input of the star. Nevertheless, our reasons for this choice are two-fold:
First, we have found that similarity solutions for the wind only exist for flows where the base density has a radial power-law slope $\lesssim -2$, thereby preventing us from using the other relations in the \citet{Hollenbach/etal/1994} model. Secondly, the $\R^{-3/2}$ scaling in \cref{eq:hollenbach_density} is consistent with XEUV disc winds \citep{Picogna/etal/2019}, which will later aid in comparing our results to \citet{Franz/etal/2020}.

\subsubsection{Ionisation front location}
\label{sec:IF_location}

The location of the ionisation front was indirectly set when we chose a constant surface inclination $\iIF$ while solving for the initial wind velocity;
now we must ensure that the gas properties within the disc are consistent with the location we have chosen. We assume that the flow in the neutral region is perpendicular to the disc mid-plane and thus the momentum equation in this region is
\begin{equation}
	\vg \frac{{\rm d} \vg}{{\rm d} z} = - \frac{\cs^2 }{\rhog}
	\frac{{\rm d} \rhog}{{\rm d} z} - \frac{ \G\Mstar z}{(\R^2 + z^2)^{3/2}},	
	\label{eq:1d_momentum}
\end{equation}
whose solution is given by: {\footnote{Note that this also describes the  plane-parallel wind solution given by \citet[][see also \citetalias{Hutchison/Laibe/Maddison/2016}]{Hutchison/Laibe/2016} but with different boundary conditions.}}
\begin{align}
	\vg & = \cs \sqrt{- \, \mathrm{W}_0 \! \! \left[ -\exp{ \left( -\frac{2 \G \Mstar}{\cs^2 \sqrt{\R^2 + z^2}} - C_1 \right) } \right]},
	\label{eq:vgas_disc}
\\
	\rhog & = \frac{\rhogO \vgO}{\vg},
	\label{eq:rhogas_disc}
\end{align}
where $\mathrm{W}_0$ is the Lambert W function. 
To find $C_1$, we look for solutions that satisfy the boundary conditions given by \cref{eq:disc_rhogmid,eq:vgn,eq:rhogn}. For an inclined ionisation front, mass conservation requires $\rhogO \vgO = \rhogn \vgn/{\rm cos} \iIF$ 
thereby fixing the initial flow velocity at the mid-plane, $\vgO$. Then, using the transcendental form of \cref{eq:vgas_disc} (evaluated at $z = 0$), the expression for $C_1$ is:
\begin{equation}
	C_1 = \MachO^2 - \ln \MachO^2 - 2 \MachK^2.
	\label{eq:const1}
\end{equation}
For convenience here and in what follows, we have defined the Mach numbers $\MachK \equiv \vK / \cs$ (where $\vK = \R \OmegaK$ is the mid-plane Keplerian velocity) and $\Mach \equiv \vg / \cs$. Using the latter definition, $\MachO$ and $\Machn$ refer to the Mach number at the mid-plane and neutral side of the ionisation front, respectively. 

We can use a similar approach to obtain an analytic relation for the height of the ionisation front, $\zIF$. At  $z = \zIF$, the transcendental form of \cref{eq:vgas_disc} gives
\begin{equation}
	\Machn^2 - \ln{\Machn^2} = \frac{2 \MachK^2}{\sqrt{1+ \left( \zIF/\R \right)^2}} + C_1.
	\label{eq:transcendental_Wi}
\end{equation}
Substituting in the expression for $C_1$ and rearranging to find $\zIF$, we obtain
\begin{equation}
	\zIF = \R \sqrt{4 \MachK^4 \left[ \Machn^2 - \MachO^2  + \ln{\left( \dfrac{\MachO^2}{\Machn^2} \right)} + 2 \MachK^2 \right]^{-2} - 1}.
	\label{eq:IF_location}
\end{equation}
Inserting the generic disc parameters from \cref{eq:disc_rhogmid,eq:disc_sigmag,eq:disc_hg,eq:disc_cs} gives an opening angle that is not constant with radius. We therefore invert \cref{eq:IF_location} to find the required disc sound speed that produces $\tan^{-1}(\zIF/\R) = \iIF$. While the deviation in $\cs$ from \cref{eq:disc_cs} is $\lesssim~2$ at all radii, the fact that $\cs$ is no longer independent of the ionisation front means that we must include this inversion calculation in our iterative scheme to find $\vgi$. Note that changing the sound speed also requires that we alter the disc scale height, which is later used in determining the dust flow. In essence, we sacrifice the simplicity of the radial power-law profiles in \cref{eq:disc_hg,eq:disc_cs} to maintain consistency with our inclined ionisation front and gas flow.

\subsection{Dust flow}
\label{sec:dust_flow}

With the gas solution fully determined up to and immediately above the ionisation front, we can now derive the corresponding solutions for the dust. We omit the dynamical effect of dust back reaction on the gas for the following reasons. In the disc, the gas pressure from the near hydrostatic equilibrium restores any perturbations caused by the dust motion. In the wind, the {\it total} dust-to-gas ratio from a realistic grain-size distribution is orders of magnitude smaller than the canonical $0.01$,
a case already shown by \citepalias{Hutchison/etal/2016} to exhibit negligible back reaction on the gas for {\it individual} grain sizes. By solving for dynamical equilibria in the $z$ direction with fixed mid-plane dust densities we are implicitly assuming that the vertical flow time (on which this equilibrium is established) is much shorter than the timescale on which the mid-plane density changes (either as a result of the wind or of radial drift in the disc mid-plane). We find that this condition is readily fulfilled in practice.

\subsubsection{Dust dynamics in the  turbulent disc}
\label{sec:disc_entrainment}

To obtain the dust velocities in the disc, we consider the momentum equation{\footnote{ The approximate equality in \cref{eq:dust_mom} is due to a small simplification we have made to the drag term, which should contain a factor of $1/(1+\eps)$ which we set to unity on the basis of the very small dust-to-gas ratios $\eps$ in our solutions.}}
\begin{equation}
    \frac{\partial \vd} {\partial t} + 
	\vd \frac{\partial \vd} {\partial z} \simeq - \frac{ \left(\vd - \vg \right)}{\ts} - \frac{\G \Mstar z}{\left( \R^2 + z^2 \right)^{3/2}},
	\label{eq:dust_mom}
\end{equation}
where $\ts$ is the stopping time in the Epstein drag regime \citep{Epstein/1924},
\begin{equation}
	\ts =  \sqrt{\frac{\pi \gamma}{8}} \frac{\rhograin s }{\cs \rhog} = \frac{\rhoeff s }{\cs \rhog},
	\label{eq:epstein drag}
\end{equation}
with $\rhograin$ the intrinsic density of individual dust grains. For convenience, we simplify the expression in the second equality by defining $\rhoeff \equiv \rhograin \sqrt{\pi \gamma/8}$ as an effective grain density.
Because we are only interested in the steady-state flow, we can omit the time derivative on the left-hand side of \cref{eq:dust_mom}. Furthermore,
for grains relevant to photoevaporation, the advection term is typically much smaller than the drag and gravitational components in the disc interior and $\vd$ in this region can be approximated by the local terminal velocity of the dust:
\begin{equation}
	 \vd = \vg - \frac{\ts \G \Mstar z}{\left( \R^2 + z^2\right)^{3/2}}.
	 \label{eq:vdust_disc}
\end{equation}

With the grain size fixed, the stopping time increases substantially along the flow on account of the exponential decline in gas density with $z$. This causes the terminal velocity approximation to break down (particularly near the ionisation front)
and failure to use the correct velocity leads to an overestimate of the dust flux leaving the disc. We emphasise that this is a general phenomenon that affects even the smallest grains due to the step-like transition at the ionisation front being comparable in width to the stopping distance of the dust. We therefore obtain $\vd$ numerically for all dust grains by solving \cref{eq:dust_mom} using Matlab's variable-step, variable-order ordinary differential equation solver ODE15s \citep{Shampine/Reichelt/1997,Shampine/Reichelt/Kierzenka/1999}. Because $\vd = 0$ is a removable singularity of \cref{eq:dust_mom}, large grains whose velocity changes signs during the flow require special attention. In these exceptional cases, we make the substitution $\sign{(\vd)}\sqrt{|w|} = \vd$ in \cref{eq:dust_mom}, which allows us to (i) remove the singularity by absorbing the multiplicative factor of $\vd$ into the derivative and (ii) maintain a real dust velocity inside of the drag term when the dust velocity is negative.

Once we have the dust velocity we proceed with finding the dust density in the disc. There is a rich body of work in the literature investigating the vertical dust distribution in turbulent discs \citep[e.g.,][\citetalias{Hutchison/Laibe/Maddison/2016}]{Takeuchi/Lin/2002,Schrapler/Henning/2004,Johansen/Klahr/2005,Fromang/Papaloizou/2006,Fromang/Nelson/2009,Charnoz/etal/2011,Birnstiel/Fang/Johansen/2016}, the vast majority of which are based on the seminal work of \citet{Dubrulle/Morfill/Sterzik/1995}. However, as pointed out by \citet{Riols/Lesur/2018}, the net vertical flow induced by disc winds can alter the dust distribution in the disc. To capture this effect in the \citet{Dubrulle/Morfill/Sterzik/1995} model, we use the dust velocity derived in \cref{eq:vdust_disc} to compute the dust density in the advection-diffusion equation,
\begin{equation}
	\frac{\partial \rhod}{\partial t}  + \frac{\partial}{\partial z} \left[ \rhod \vd - \rhog \Dd  \frac{\partial}{\partial z} \left( \frac{\rhod}{\rhog}\right) \right] = 0,
	\label{eq:dust_adv_diff}
\end{equation}
where we approximate the diffusion coefficient for the dust, $\Dd$, using the expression from \citet{Charnoz/etal/2011}:
\begin{equation}
	\Dd \approx \frac{\alpha \cs \Hg}{\Sc}.
	\label{eq:small_kappa_t}
\end{equation}
Here $\alpha$ is the Shakura-Sunyaev turbulence parameter \citep{Shakura/Sunyaev/1973} and $\Sc$ is the dimensionless Schmidt number. There are a number of definitions for $\Sc$ in the literature \citep[see][for a discussion]{Youdin/Lithwick/2007,Laibe/2014}, but we have opted to use $\Sc = 1 + \St$ since we are only considering motion in the vertical direction \citep{Laibe/2014}.

A shortcoming of this model is that it assumes uniform turbulence in the disc, which may in reality vary depending on the physical mechanism generating the turbulence. For example, \citet{Shi/Chiang/2014} show that gravito-turbulence is relatively uniform vertically (only differing by a factor of $\unsim 2$ from mid-plane to surface), but turbulence induced by the magnetorotational instability is more variable (fluctuating by a factor of $\unsim 15$). Another shortcoming is that it assumes that the Stokes number ($\St \equiv \OmegaK \ts$) is much smaller than unity. In practice we find that all grains that enter the wind satisfy this condition in the turbulent region of the disc, but not necessarily in the wind. In \cref{sec:preliminaries} we discuss some limitations to our model that arise when the stopping time becomes comparable with the timescale on which
dust crosses the ionisation front. 
 
A significant issue in coupling the wind to a turbulent disc is that we need to specify the value of $\Dd$ in the wind (i.e. above the ionisation front).
In default of a model for turbulence in the wind we will assume that
this region of the flow is laminar so that $\Dd$ tends to zero. While it
is reasonable to assume that the physical scale on which the disc
makes the transition from a neutral to ionised state is governed by
recombination physics 
the relevant length scale on which $\Dd$ declines at the
ionisation front is not obvious. 
Clearly, step functions for $\Dd$ and $\rhog$ should be avoided because the derivatives in \cref{eq:dust_adv_diff} would be dependent on the resolution of our numerical grid.
For this reason we smooth the disc quantities $\Dd$, $\ts$, $\cs$, and $\vg$ (smoothing of $\rhog$ is indirectly set by $\vg$ through mass conservation) onto their respective values in the wind
    \footnote{Because the solutions in this section are strictly 1D, we map the 2D gas streamline onto the vertical coordinate assuming $z$ is equal to the distance along the streamline. In so doing we inherently assume there are no radial changes to the gravitational force and that the dust follows the same trajectory as the gas, both of which are approximately true near the base of the wind.}
using a tanh function parameterised in terms of a length scale $W$. For example, the smoothed functional form for the Stokes number is given by
\begin{equation}
    \St(z) = \frac{1}{2} \left[ (\tilde{\St}(z) + \Sti) - (\tilde{\St}(z) -\Sti)
    \tanh{\left( \frac{z - \zIF}{W/3} \right)} \right]
    \label{eq:tildest}
\end{equation}
where $\tilde{\St}(z)$ is obtained from \cref{eq:epstein drag} using non-smoothed disc variables and $\Sti$ is the value of the Stokes number in the ionised flow above the front. Throughout the discussion of the results we characterise the dust properties in terms of $\St(\zIF)$; since the Stokes number increases strongly as the density drops at the front, $\St(\zIF)$ can be seen to be around half of the Stokes number in the fully ionised wind above the ionisation front.
By applying these smoothed forms for variables at the ionisation front we can test the sensitivity of
our results to the assumptions we have made about the relevant
length scales while ensuring that $W$ is sufficiently above the grid scale such that spatial derivatives can be reliably computed. When not otherwise specified, we will estimate the width by
\begin{equation}
    W = \frac{\mH}{\sigma \rhog(\zIF)},
    \label{eq:width}
\end{equation}
where $\sigma = 6.3 \times 10^{-18} \cm^2$ is the photoionisation cross-section for neutral hydrogen and $\rhog(\zIF)$ is the smoothed density at the ionisation front.

The conceptually simplest approach is to evolve \cref{eq:dust_adv_diff} on an Eulerian grid for a specific grain size in a background gas velocity and density field that smoothly connects onto the wind.
In this implementation we need only to fix the dust density at the mid-plane; an upper boundary condition is unnecessary as long as the grid
extends into the region where the flow becomes laminar and 
\cref{eq:dust_adv_diff} effectively becomes first order. In practice, as the solution evolves towards a steady state, it `finds' a steady-state flux:
\begin{equation}
	\F=-{\Dd}\biggl(\frac{\mathrm{d} \rhod}{\mathrm{d} z} - \frac{\mathrm{d} \ln{\rhog}}{\mathrm{d} z}  \rhod \biggr) + \rhod \vd,
	\label{eq:dust_adv_diff_steady_state}
\end{equation}
and corresponding gradient of the dust-to-gas ratio at the mid-plane with a smoothly varying steady-state dust density profile. 

With the lack of a second fully-prescribed boundary condition, it is not immediately obvious as to why the Eulerian simulation prefers any one solution over another. Fruitfully, we can investigate the family of analytic solutions to \cref{eq:dust_adv_diff} in a steady state characterised by constant $\F$ (or equivalently by choosing the gradient of the dust-to-gas ratio at $z=0$).
Assuming $\Dd \neq 0$, these solutions take the general form
\begin{equation}
	\rhod = \eps \rhog e^{I} \left( 1 - \int^z_0 \dfrac{\F e^{-I}}{\eps \rhog \Dd}  \mathrm{d} z' \right),
	\label{eq:rhodust_disc}
\end{equation}
where the mid-plane dust-to-gas ratio $\eps$ is a function of grain size and, for convenience, we have defined
\begin{equation}
    I = \int^z_0 \frac{\vd}{\Dd} \mathrm{d} z'.
\end{equation}
After experimenting with a range of 
values of $\F$ it is readily apparent that $\F$ only affects the density profile in the region near the ionisation front itself, specifically where $\Dd$ is close to zero. Small values of $\F$ cause the density to blow up to $\infty$ while large values cause the density to plummet to $-\infty$. The explanation for this behaviour is as follows. Since $\F$ effectively sets the total dust flux through the disc, selecting an arbitrarily large flux implies that a correspondingly large amount of dust is being lost to the wind. With the advective velocity fixed, the only way to impose such a condition is to adjust the dust profile so as deliver this flux; in practice this involves  
establishing a large negative or positive gradient in the dust-to-gas ratio in that vicinity.
These requirements cause the density to become unbounded at the ionisation front for all but an extremely narrow range of fluxes that only differ beyond their six or seventh significant digit -- for all intents and purposes, a single value of $\F$.

With this insight it can be seen that the value of $\F$ selected
is such that the disc and wind solutions connect smoothly at the ionisation front (i.e. no maxima, minima, or kinks). Thus it comes as no surprise that this `unique' value of $\F$ reproduces the steady-state solution obtained by evolving \cref{eq:dust_adv_diff} numerically.
We therefore conclude that our selection of $\F$ in the analytic model is both unique and physically motivated. The obvious benefit to using \cref{eq:rhodust_disc} over conventional numerical schemes is that, even after sweeping over possible $\F$ values, we can obtain the steady-state dust density in seconds -- a speed-up of $\unsim10^4$ times over the numerical method detailed below. Furthermore, the analytic solution scales well to very high resolutions, requiring only a few minutes to compute the vertical dust velocity and density at a given disc radius on a grid with 10 million points. This latter quality is particularly important in this study so that we can ensure that the ionisation front is always adequately resolved as we lower $W$ (the smallest of which requires 60 million points).

As a final remark, it is important to note that \cref{eq:rhodust_disc} breaks down as $\Dd \to 0$, which, due to the tanh smoothing, always occurs within $\unsim 3W$ of the ionisation front, regardless of the underlying disc parameters. Likewise, when $\alpha \lesssim 10^{-3}$  \cref{eq:rhodust_disc} can also predict unphysical behaviour, even in the bulk of the disc. In these regions where the turbulent diffusion is small, we can (in most cases) circumvent these issues by transitioning to a purely advective solution where the two solutions smoothly connect. In practice we do this by calculating the derivative of $\rhod = \dot{m_{\rm d}}/\vd$ for each potential transition point, where $\dot{m_{\rm d}}$ is a constant representing the mass flux at each of these points. We start at the last defined point in the diffusive solution and work backwards toward the disc mid-plane until we find the point where the slope from the two solutions are equal. We have found that the above procedure agrees well with numerical simulations (as long as sufficient resolution is used), thus extending the parameter space that we can model.

\subsection{Comparing the semi-analytic model to time-dependent calculations}
\label{sec:testing}
%
\begin{figure*}
	\centering{\includegraphics[width=\textwidth]{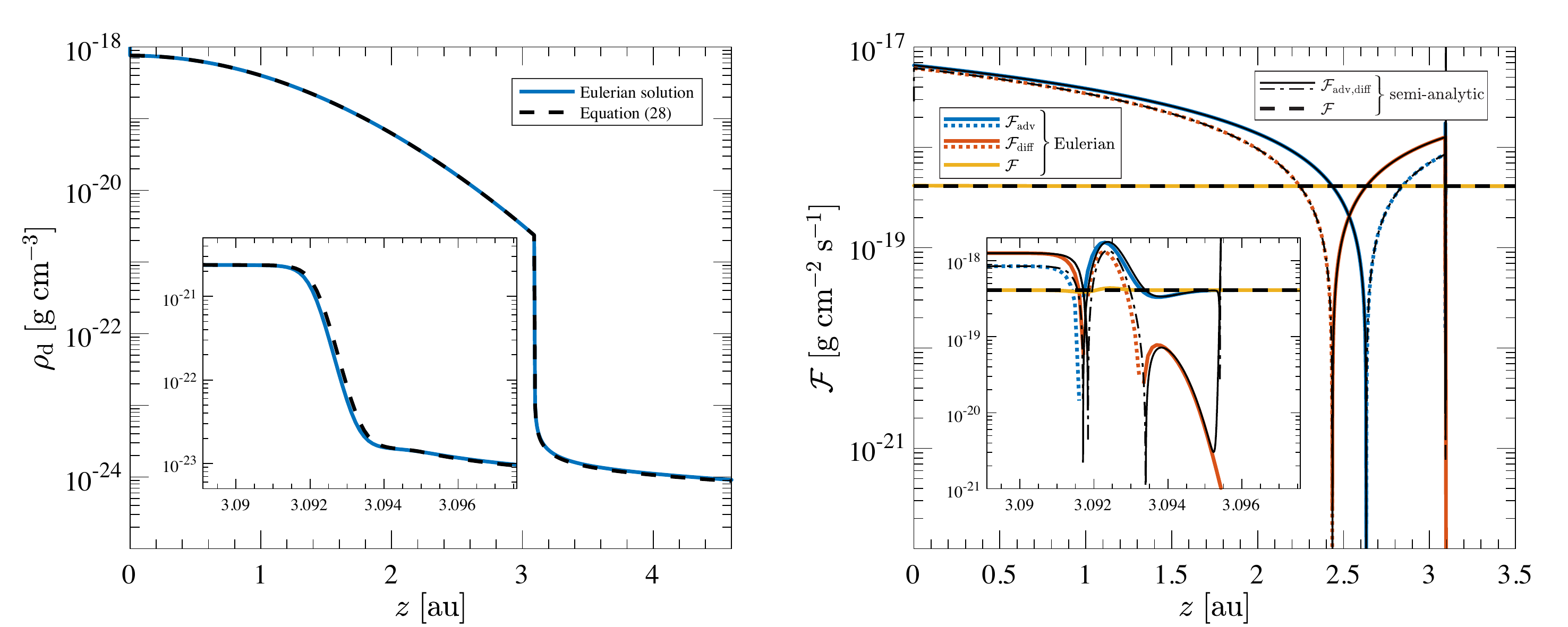}}
	\caption{Comparison between our semi-analytic model (black lines) and the steady-state hydrodynamical Eulerian solution (coloured lines) for dust grains with size $s = 0.811\mum$ at $\R = 10\au$. {\it Left:} the dust density as a function of height above the mid-plane. The inset panel is a zoom-in of the region $z \in [\zIF\pm 3W]$, where $\zIF = 3.09336\au$, and shows that the ionisation front is well-resolved. {\it Right:} the corresponding advective, diffusive, and total dust flux as a function of $z$. Negative fluxes are drawn with dotted lines in the Eulerian solution and dot-dashed lines in the semi-analytic solution. The inset panel again zooms in on the ionisation front and shows the complex behaviour of the advective and diffusive fluxes near $\zIF$. Resolving this region is crucial to finding the correct dust flux leaving the disc.}
	\label{fig:numerical_test}
\end{figure*}
To assess the validity of our disc model, we solved the full time-dependent hydrodynamic \cref{eq:dust_mom,eq:dust_adv_diff} using a modified version of the grid code described in Appendix A2 of \citet{Hutchison/Price/Laibe/2018}. 
Assuming $\Phii = \Phi_{45}$ and $\R = 10\au$, we discretise the region $z \in [0,\,1.5\zIF]$ with $N_z = 40~002$ cell-centred grid points, including ghost points, such that we span the width of the ionisation front $W \approx 0.002\au$ with $\unsim 17$ grid points. We specify the inner boundary conditions at $z = 0$ to be ${\rm d}\vg/{\rm d}z = 0$ and $\rhog = \rhogO$ while leaving the outer boundary conditions at $z = \zIF$ open (i.e. discretised versions of \cref{eq:1d_momentum,eq:dust_mom} appropriate for describing the mid-point between adjacent grid points $N_z-1$ and $N_z$). We directly import the gas velocity, gas density, and diffusion coefficient calculated from our model (including the smoothing between disc and wind), but start the dust from rest ($\vd = 0$) with a hydrostatic density profile that extends smoothly to the outer edge of our numerical grid. We then allow the velocity and density to evolve in time until dynamic equilibrium is reached.

The coloured lines in \cref{fig:numerical_test} show the steady-state Eulerian density and fluxes for a model with $\alpha = 0.05$, dust with size $s = 0.811\mum$, intrinsic grain density $\rhograin = 3 \gram \cm^{-3}$, and a mid-plane dust-to-gas ratio per local logarithmic size bin of $\eps = 2.53\times 10^{-6}$
(consistent with a MRN grain-size distribution spanning $s \in [1 \nm,\, 10\cm]$, discretised into $100$ equally-spaced logarithmic bins, and an over all mid-plane dust-to-gas ratio of $0.01${\footnote{Note that the normalisation of the dust density does not affect the form of the resulting profiles, given the neglect of back reaction of dust on the gas.}}). Corresponding results from our semi-analytic model calculated on a grid with $N_z = 10^6$ points are overlaid in black and show excellent agreement with the hydrodynamic solution. Note that the grain size shown in this example is such that the dust velocity changes sign below
the ionisation front and the net upward flux of dust at the front is driven by diffusion. Small deviations can be seen in the transition region near the ionisation front (see inset panels), but are most likely caused by the different resolutions used for each model (the higher resolution in the analytic case helps to find a better fit for the purely advective solution in the wind). Even with the difference in resolution, the semi-analytic solution takes seconds to compute while the steady-state hydrodynamic solution takes days to reach a true dynamic equilibrium.

\section{Results}
\label{sec:results}

We start by defining a few quantities/regimes that will aid in the analysis of our results. We then present a suite of simulations that vary grain size, turbulence, and width of the ionisation front at a single radial location ($\R = 10\au$) and mid-plane gas density ($\Sigmag = 10 \gram \cm^{-2}$). We examine the possible dust trajectories allowed by the wind at this location and compare the range of delivered grains to the ionisation front to the maximum entrainable grain size by the wind. Finally, we show how the grain properties change with disc radius and ionising flux. Unless otherwise specified, we assume the following set of fiducial parameters: a solar-mass star with an EUV luminosity of $\Phi_{45}$ and an intrinsic grain density of $3 \gram \cm^{-3}$.

\subsection{Preliminaries}
\label{sec:preliminaries}

%
\begin{figure*}
	\centering{\includegraphics[width=\textwidth]{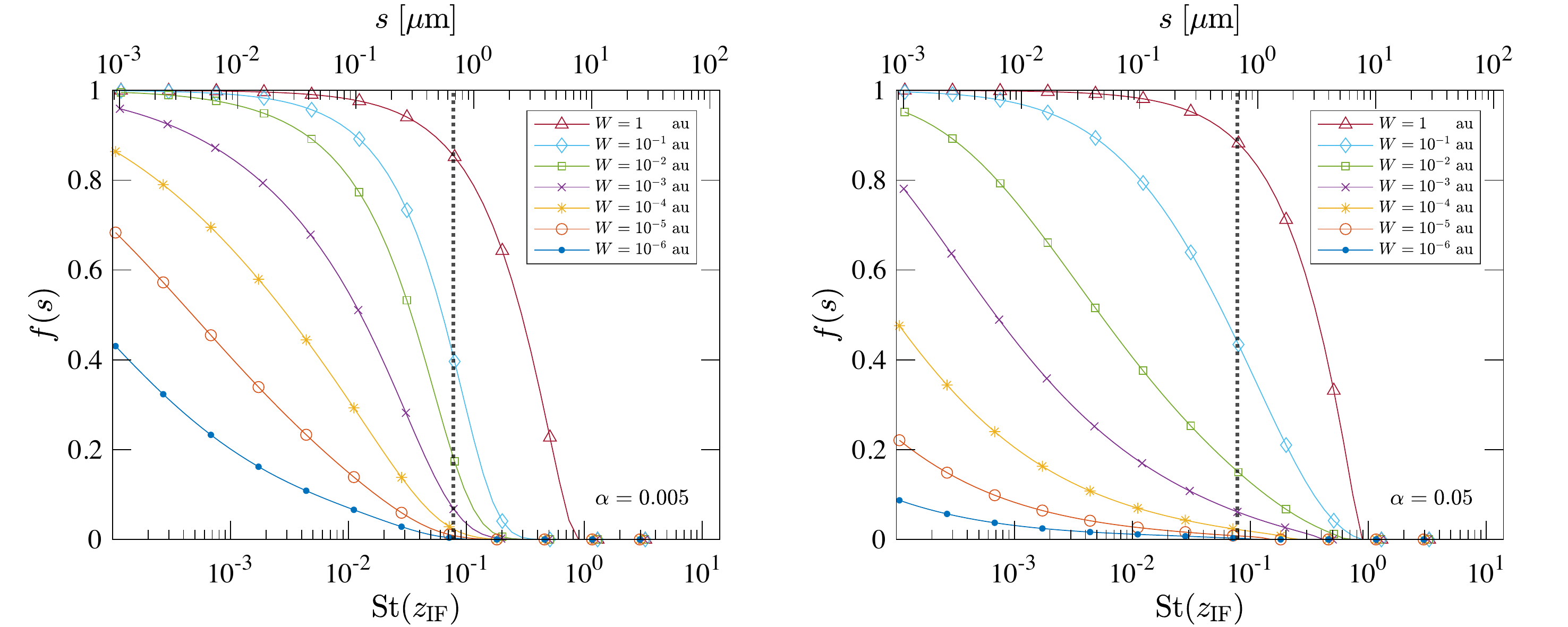}}
	\caption{The normalised flux (ratio of the flux of given grain size to its value if advected with the gas) as a function of grain size (upper horizontal axis) and Stokes number at the ionisation front (lower horizontal axis). Each line denotes the relation for fixed value of the width, $W$, of the ionisation front (inset legend). The vertical dashed line indicates the critical grain size ($\scrit$) or Stokes number ($\Stcrit$) for which the dust velocity is zero at the ionisation front. The left and right panels represent $\alpha = 0.005$ and $0.05$, respectively. Note that the conversion between Stokes number and grain size is not constant with $W$, but variations for the cases shown are small (e.g. the left edge of each curve corresponds to $s = 10^{-3}\mum$).}
	\label{fig:f_ratio_plots}
\end{figure*}

\begin{figure*}
	\centering{\includegraphics[width=\textwidth]{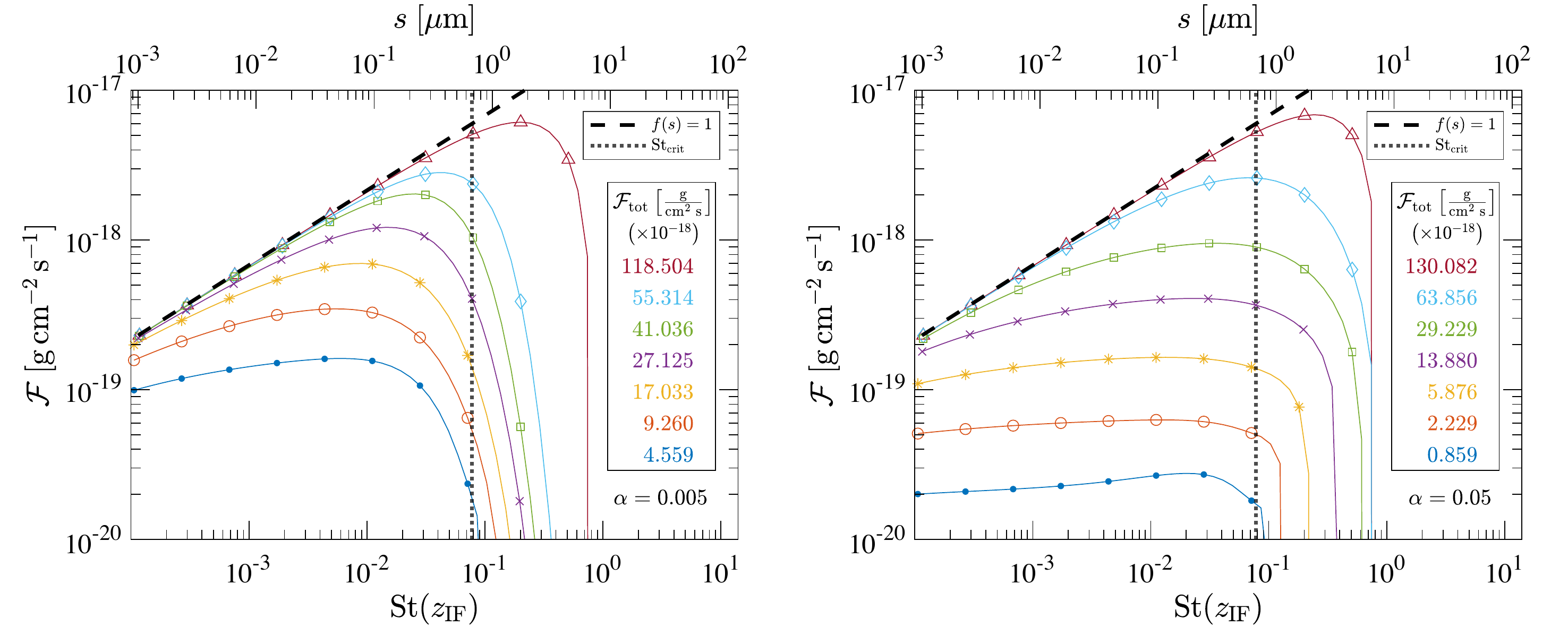}}
	\caption{The flux of dust per logarithmic grain size interval for the case of dust following an MRN size distribution in the disc mid-plane. The curves and vertical dotted line have the same meaning as in \cref{fig:f_ratio_plots} while the oblique dashed line indicates the flux value in the advection limit ($f=1$). The total mass flux that escapes to the wind at this radius ($\R =10\au$) is listed for each curve in the inset box on the right of each figure, assuming a maximum grain size of $\smaxmrn = 10\cm$ and a total mid-plane dust-to-gas ratio of $0.01$.}
	\label{fig:F_St_limits}
\end{figure*}

Much of the analysis and discussion that follows is framed in terms of the dimensionless Stokes number because it both highlights the important physics of the problem and is relatively unaffected by changes to the ionising luminosity. On the other hand, $\St$ depends on the background gas density and varies substantially from the disc mid-plane to the wind. To avoid this latter ambiguity our discussion of $\St$ is confined almost exclusively to the the ionisation front, particularly the smoothed value defined in \cref{eq:tildest}. These values can be related back to physical grain sizes by using the secondary axes provided in \cref{fig:f_ratio_plots,fig:F_St_limits,fig:Stmax_Stcrit}.
 
To help with our analysis, we define the normalised flux
\begin{equation}
    \f \equiv \frac{\F}{\Fadv} = 1 + \frac{\Fdiff}{\Fadv},
    \label{eq:normalised_flux}
\end{equation}
where $\Fadv$ and $\Fdiff$ are the advective and diffusive components of the total flux. Importantly, $\f$ is effectively bound between $[0,1]$ under realistic conditions.
The upper bound $\f = 1$ may correspond to the case that the dust and gas kinematics are the same (i.e. small grains with small $\ts$) so that the dust-to-gas ratio for that grain size remains constant along the flow trajectory.
However $\f = 1$ could also correspond to a purely advective solution (i.e. when diffusion is negligible) but with the dust and gas velocity diverging with increasing $z$ and the dust-to-gas ratio varying so as to maintain constant flux (provided that the dust velocity remains positive at all $z$). Either way, we will refer to the case $\f = 1$ as the {\it advection limit}.

In the opposite extreme of $\f \to 0$, we approach what we call the {\it diffusive limit}. Here $\Fdiff \sim -\Fadv$ and, similar to the advective limit, can correspond to different scenarios. 
The first scenario is relevant to dust grains that are large enough to decouple from the gas flow in the disc such that the dust velocity becomes negative.
The resulting sign reversal of $\Fadv$ first occurs at the ionisation front for a critical grain size $\scrit$ (or in terms of Stokes numbers $\Stcrit$).
In a purely advective flow, grains with $s > \scrit$ would experience a runaway pile-up before ever reaching the ionisation front. However, diffusion considerably smooths the advective pile-ups in these situations and takes over as the sole delivery mechanism to the ionisation front.
A second scenario occurs when $\Fadv$ is still positive at the ionisation front, but is either weak (e.g. due to large $\St$) or the turbulent diffusivity is strong (e.g. due to a steep gradient in the dust-to-gas ratio or large $\alpha$).
The fact that the dust-to-gas ratio increases with $z$ for the purely advective solution means that the diffusive flux tends to be negative and $\F < \Fadv$. In other words, diffusion actually opposes the delivery of dust to the ionisation front. In either scenario, we see the diffusion limit is characterised by a low normalised flux $\f$.

More quantitatively, we can obtain a limiting form for the diffusive limit by considering the inability of dust to accelerate as rapidly as the gas over the finite width of the ionisation front. As the gas accelerates to $\vgi$ across width $W$, the dust is accelerated by $\Delta \vdi \sim \vgi \delta t/\ts$, where $\delta t \sim W/\Delta \vdi$ is the timescale for dust to cross the front. Thus, in this limit, $\Delta \vdi \sim \sqrt{\vgi W/\ts}$; in the absence of diffusion, this would mean that the dust-to-gas ratio was boosted by a ratio $\frac{\vgi}{\Delta \vdi}$ across the front in order to satisfy continuity for the dust. However {\it if} diffusion is strong enough to iron out such a strong increase in dust-to-gas ratio across the front, this limits the value of $\f$ to:
\begin{equation}
    \f_{\rm diff} \sim \frac{\Delta \vdi}{\vgi} = \sqrt{\frac{W \OmegaK}{\StIF \vgi}}.
    \label{eq:diff_limit}
\end{equation}
Interestingly, the solutions can approach this limit without imposing strong turbulence in the disc, provided that the front is sufficiently thin. This is because the sluggish acceleration of the dust produces a steep positive gradient in the dust-to-gas ratio at the ionisation front, which in turn excites strong diffusive mixing back into the disc.

It is however worth raising a caveat about the reality of the solutions in the diffusion limit which we cannot
investigate further within the framework of the \citet{Dubrulle/Morfill/Sterzik/1995} formulation. When we solve the advection diffusion equation
(\crefalt{eq:dust_adv_diff}), in conjunction with \cref{eq:dust_mom} we are assuming that, whereas the Stokes number controls the coupling of the grain motion to the {\it mean} flow, the {\it turbulent} motion of the grains is equal to that prescribed for the gas. Although we allow the Schmidt number (\crefalt{eq:small_kappa_t}) to vary with Stokes number as $1+\St$, this means that since $\St < 1$ for grains entering the wind, in practice $\Sc \sim 1$ and thus it is assumed that the grains experience the same diffusivity as the gas. 
This is a reasonable assumption for many applications where the relevant timescale associated with the turbulence is the local dynamical time,
$\Omega^{-1}$, but in the present case we are concerned with the value of the diffusivity over a very small spatial region through which the grains are passing on a much shorter timescale. We thus caution that we are likely to be over-estimating the effective diffusivity of the dust at the front. If the effective diffusivity should indeed be lower, we can expect the ironing out of small scale variations in the dust-to-gas ratio at the front to be less severe (i.e. the process that is responsible for driving the solution to  the diffusion limit). Exploring this issue is beyond the scope of the present paper and would require an explicit modeling of dust particle motions that are only partially coupled to the turbulent gaseous background.

\subsection{Delivering dust to the ionisation front}
\label{sec:dust_delivery}

%

\cref{fig:f_ratio_plots} displays the normalised flux $f(s)$ defined in the previous section for the cases $\alpha = [0.005,\,0.05]$ and a range of values of $W$, the scale length over which the diffusivity and gas density are smoothed at the ionisation front. For a fixed gas profile, the grain size (upper horizontal axis) scales linearly with the Stokes number at the ionisation front, $\StIF$ (lower horizontal axis), the smoothed value at $z = \zIF$ (\crefalt{eq:tildest}); the vertical dotted line represents $\Stcrit$ (the corresponding grain size is denoted $\scrit$), the maximum value of $\StIF$ for which the dust velocity is upwardly directed for all $z < \zIF$ 
in the limit of an infinitely thin ionisation front. Deviations from this infinitely thin limit are small as long as $W \ll \zIF$, a condition we expect to hold for real systems (e.g. \crefalt{eq:width}). However, larger widths can experience critical sizes that are a factor of a few larger 
(e.g. $\scrit \simeq 2 \mum$ for $W = 1\au$ at $\R = 10 \au$, a factor of $\unsim2.5$ higher than our fiducial case).
\cref{fig:f_ratio_plots} shows that for large $W$ and low $\alpha$, $f(s)$ is close to the advection limit and declines steeply for $\StIF > \Stcrit$. In line with our earlier prediction, as diffusion increases in importance (lower $W$ and higher $\alpha$), the ironing out of the positive gradient in the dust-to-gas ratio at the ionisation front acts to suppress the flow of dust.
At the same time diffusion facilitates dust transport across the front for grains with $\StIF \gtrsim \Stcrit$, with the upper limit on grain size increasing weakly with $W$. We will use $\slimit$ and $\Stlimit$ to distinguish this upper size limit set by the existence of solutions with positive flux from the maximum entrainable size limit set by the wind (see \cref{sec:wind_solution}).

We provide a quantitative demonstration of the way that the fluxes vary between the advection and diffusive limits in the following section but, for now, note qualitatively how this affects the behaviour seen in \cref{fig:f_ratio_plots}.
For high $W$, dust with $\StIF < \Stcrit$ has $\f \sim 1$ (efficient delivery to the ionisation front).
In the limit of low $W$, grains are not efficiently delivered even for $\StIF \ll \Stcrit$ because in the diffusive limit $\f \propto 1/\sqrt{\StIF}$.

In practice, therefore, the value of $W$ influences whether grains with $\StIF$ up to a few times $\Stcrit$ (i.e. of size $s \gtrsim \scrit =  0.8 \mum$ for these 
parameters) can reach the wind. It can be seen from \cref{fig:f_ratio_plots}
that the decline in $\f$ as $W$ is decreased is stronger for higher $\alpha$ because this means that fluxes reach the diffusive limit at higher $W$. Thus the ability of dust to reach the wind is jointly
controlled by the width of the acceleration region and also by whether there are microphysical processes at the front that can mix dust across this region.

\subsection{Dust delivery for a MRN dust size distribution.}

We now consider the case that the dust in the mid-plane follows a MRN distribution (number of grains in size range $s$ to $s+ds$ scaling as $s^{-3.5} ds$ up to a maximum grain size $\smaxmrn$). Assuming spherical dust grains with uniform intrinsic density, the mid-plane dust-to-gas ratio per logarithmic size bin, $\eps$, scales as $\sqrt{s}$.
\cref{fig:F_St_limits} depicts the flux of dust per logarithmic size interval ($\F$, i.e. \crefalt{eq:dust_adv_diff_steady_state}) as a function of grain size (upper scale) and $\StIF$
(lower scale) when $\alpha = [0.005,\,0.05]$.
The heavy dashed line represents the advection limit (i.e. where all grains with $\StIF < \Stcrit$ are advected
with a flux that is the product of the gas flux and the mid-plane dust-to-gas ratio for each size bin: $\F = \eps \vgO \rhogO \propto \sqrt{s}$).
The main utility of this plot is that it shows (i) how far below the maximum (advection) limit the dust flux falls and (ii) the grain size distribution of the dust delivered to the ionisation front. In each panel the various model lines correspond to the same values of $W$ in \cref{fig:f_ratio_plots}.

\cref{fig:F_St_limits} confirms the behaviour described in \cref{sec:dust_delivery} in that, in the limit of large $W$, the dust loss is close to the advection limit and then rolls over at a grain size corresponding to a few times $\Stcrit$ ($\unsim1\mum$ for $\Phi_{45}$ and $\R = 10\au$).
As $W$ is reduced, the mass loss decreases across all grain sizes, but with a greater reduction for larger grains.
The limiting behaviour at small $W$ is that the mass loss per logarithmic size bin is roughly constant for grains up to $\Stcrit$ and then cuts off abruptly for larger grain sizes.
The behaviour is similar for the two $\alpha$ values but the $\alpha = 0.05$ profiles start to deviate from the advection limit at higher $W$ values, due to more efficient mixing across the ionisation front.

\begin{figure*}
	\centering{\includegraphics[width=\textwidth]{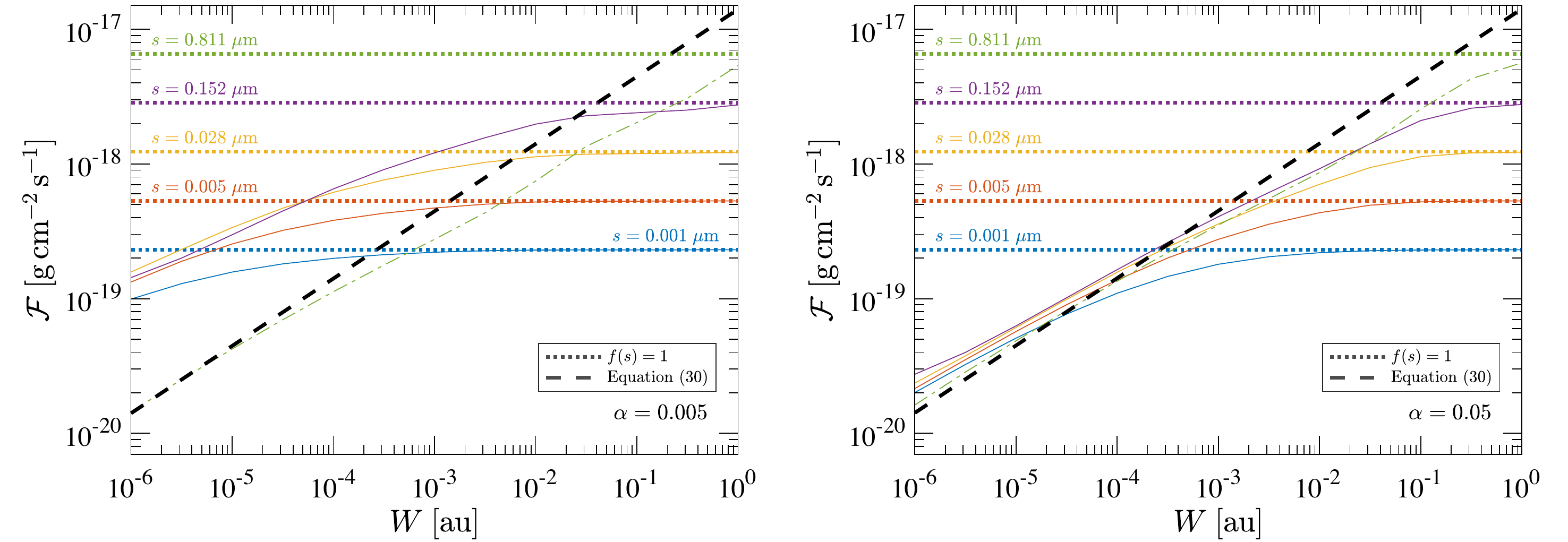}}
	\caption{The flux of dust per logarithmic grain size interval for a variety of grain sizes as a function of the ionisation front width $W$. For each grain size, the flux value corresponding to the advection limit ($f=1$) is shown by a horizontal dotted line of the same colour. The oblique dashed line indicates the diffusion limit (\crefalt{eq:diff_limit}).
	As expected, the flux converges to the diffusive limit at small $W$ and large $\alpha$ (right panel). For smaller $\alpha$ (left panel), the flux at small $W$ is caught in an intermediate regime that is strongly influenced by both advection and diffusion. A notable exception is $s = 0.811 \mum$ (dot-dashed line), which converges to the diffusive limit regardless of $\alpha$ because it has a Stokes number larger than $\Stcrit$ (i.e. no advective flux at the ionisation front).}
	\label{fig:F_W_limits}
\end{figure*}
In \cref{fig:F_W_limits}, $\F$ is plotted as a function of $W$ for five different grain sizes. For each size, a horizontal dotted line marks the corresponding advection limit.
Again it is evident that $\F$ tends to the advection limit at high $W$ and that the $W$ value at which this transition occurs increases for higher $\alpha$. In the $\alpha = 0.05$ case, all grain sizes converge to the same relation at low $W$.
Noting that in the diffusion limit (\crefalt{eq:diff_limit}), $\f(s) \propto 1/\sqrt{\StIF} \propto 1/\sqrt{s}$ and that, for a MRN distribution, $\F \propto \sqrt{s} \f$, it follows that $\F$ is independent of $s$ in the diffusion limit and (from \crefalt{eq:diff_limit}), scales as $\sqrt{W}$.
The convergence of $\F$ onto this locus (shown as the dashed line) at low $W$ is thus confirmation of the way that diffusion limits the dust flux for $\StIF < \Stcrit$ and enhances the dust flux for $\StIF > \Stcrit$ (e.g. $s = 0.811 \mum$, distinguished from the others by a dot-dashed line).
For $\alpha = 0.005$, we see qualitatively similar behaviour in that $\F$ falls below the advection limit as $W$ is decreased, but has not converged to the diffusion limit for the values of $W$ shown. 
The behaviour reported here, where diffusion can in some cases suppress the dust loss in the wind is a consequence of solving \cref{eq:dust_adv_diff} \citep{Dubrulle/Morfill/Sterzik/1995} in the case where diffusion is effective across the narrow ionisation front; we however draw attention to the discussion at the end of
\cref{sec:preliminaries} concerning the realism of such solutions in cases where the dust is unable to couple to turbulent motions in the gas on the length scale of the front.

\subsection{Connecting to the wind solution}
\label{sec:wind_solution}

%
\begin{figure*}
	\centering{\includegraphics[width=\textwidth]{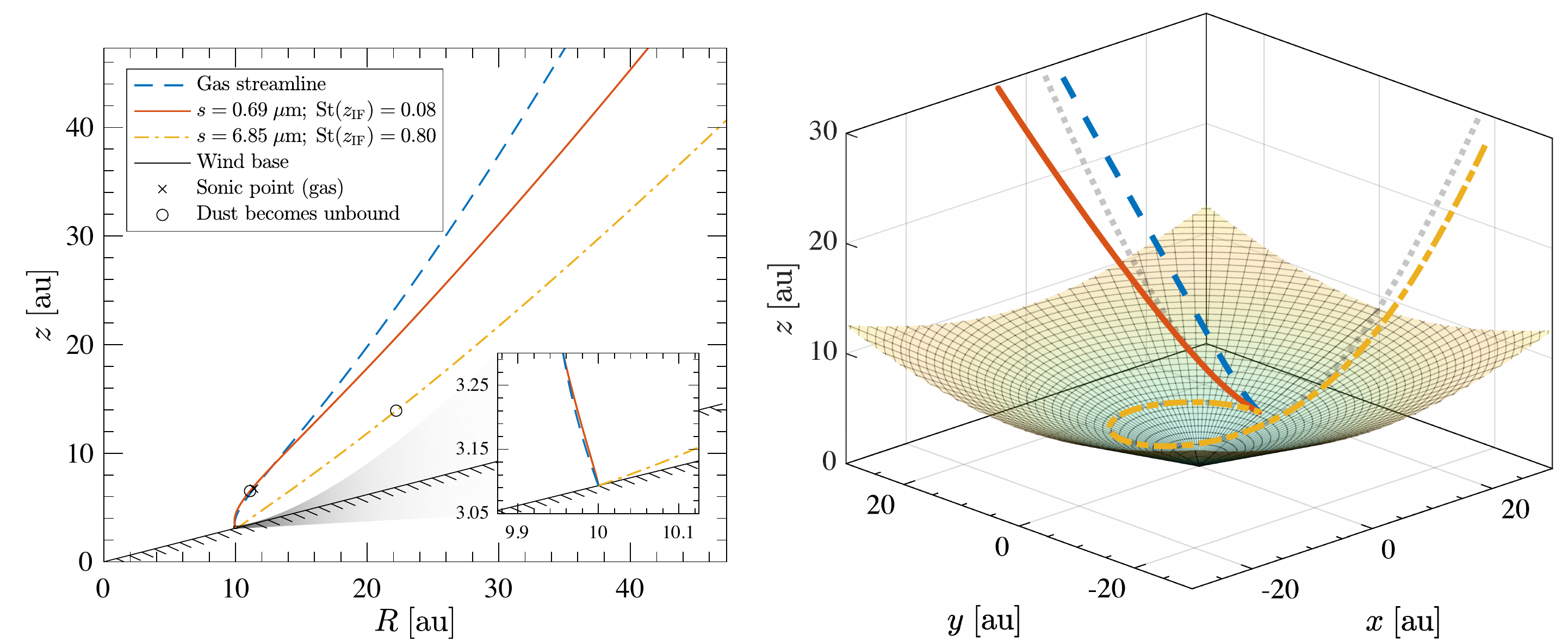}}
	\caption{{\bf Left}: Gas streamline (blue dashed line) and trajectories for two dust sizes at $\R = 10 \au$. The orange solid line shows the trajectory of a grain near $\Stcrit \sim 0.08$ while the yellow dot-dashed line is the trajectory for the maximum entrainable grain size $\Stmax \sim 0.8$. The cross marks the sonic point for the gas and the open circles where the dust grains become unbound. The inset panel shows a square zoomed-in region of width $0.25 \au$ near the base of the wind (black dotted line). Grains larger than $\Stmax$ immediately intersect with the disc rather than raining down at larger radii.
	{\bf Right}: The corresponding 3D streamline/trajectories for the gas and dust in the left panel. The grey dotted lines show how the dust trajectories would change if the azimuthal drag were neglected.}
	\label{fig:dust_streamlines}
\end{figure*}

We examine the trajectories of dust grains that reach the ionisation front by considering their entrainment in an ionised wind whose structure and kinematics is given by a variant of the self-similar wind solution of \citet{Clarke/Alexander/2016}. Specifically, whereas \citet{Clarke/Alexander/2016} considered pressure driven winds that launch from the disc mid-plane, here we consider winds that launch from the finite height above the disc at which ionisation balance is achieved. In order to preserve the assumption of self-similarity it is necessary to describe the launching surface as an axisymmetric inclined plane, specified by the angle
$\iIF$ between its normal and the normal to the disc mid-plane. Here we adopt $\iIF = 0.3$ and find that the maximum launch velocity  of gas above the ionisation front which allows the flow to make a smooth transition between subsonic and supersonic flow is $0.44 \csi = 4.26 \kms$. Above the ionisation front we consider the dynamical evolution of dust grains that are subject to the combination of acceleration due to gravitational,
centrifugal, and drag forces. We do not assume that the grains are always traveling at their local terminal velocity (\crefalt{eq:vdust_disc}) as the lower densities in the wind do not necessarily  allow application of this `short friction time' assumption. The Lagrangian equation of motion in the $(\R,z,\phi)$ directions can be written as:
\begin{align}
    \frac{D v_{\rm \R}}{D t} &= - \frac{v_{\R}-u_{\R}}{\ts}  - \frac{\G \Mstar
\R}{\left(\R^2 + z^2 \right)^{3/2}} +
    \frac{l_\phi^2}{\R^3},
    \label{eq:fR}
\\
    \frac{D l_\phi}{D t} &= - \frac{l_\phi-\R u_\phi}{\ts} 
    \label{eq:fphi},
\\
    \frac{D v_{\rm z}}{D t} &= - \frac{v_z-u_z}{\ts}  - \frac{\G \Mstar z}{\left(\R^2 + z^2 \right)^{3/2}},
    \label{eq:fz}
\end{align}
where $l_\phi$ is the specific angular momentum of the dust. It is
assumed that each gas streamline is characterised by its specific angular momentum at the wind base.
 
As described in \cref{sec:ui}, the gas emerges nearly perpendicularly to the ionisation front and thus at an angle $\iIF$ to the vertical. 
The gas velocity at the base of the flow is a significant
fraction of the sound speed in the ionised gas while the dust crosses the ionisation front with a speed close to the gas flow below the front (which is subsonic with respect to the cool neutral disc gas). Consequently the initial drag terms acting on the dust just above the front are close to those experienced by a stationary grain. Assuming $v_{\R} = u_{\R} = v_z = 0$, the ratio of the vertical and radial components of the momentum equation is given by
\begin{equation}
    \frac{D v_z}{D v_{\R}} = \frac{\sec^3{\iIF} - \chi \tan{\iIF}} {\chi(\sec^3{\iIF} - 1)},
    \label{eqn:force_ratio}
\end{equation}
where
\begin{equation}
    \chi \equiv \frac{\G \Mstar}{\R^2}\frac{t_s}{|\vg|} = \St \frac{\vK}{|\vg|}.
    \label{eqn:chi_def}
\end{equation}
If $D v_z/D v_{\R} < \tan{\iIF}$, or equivalently $\chi > \cot{\iIF}$, grains promptly re-intersect the ionisation front and are not entrained. For $R = 10 \au$ and $\iIF = 0.3$, this would imply that the maximum Stokes number for which grains can be entrained in the wind occurs a little above unity ($\Sti = \Stimax \sim 1.47$, corresponding to a physical size of approximately $0.07$ and $7 \mum$ for $\Phi_{41}$ and $\Phi_{45}$, respectively). However, in the context of our smoothed ionisation front, we must scale this value by $\unsim 1/2$ (since the gas velocity is approximately half of the ionised wind velocity) in order to be consistent with the Stokes numbers we measure at $\zIF$ in our model. Thus, in everything that follows, we use $\Stmax$ to refer to the scaled maximum measured at the mid-point between the disc and wind. Importantly, the flux delivered to the ionisation front always cuts off below this limit, only approaching $\Stmax$ at very large $W$ (see \cref{fig:f_ratio_plots,fig:F_St_limits}). We therefore expect all grains that pass through the ionisation front at $R = 10 \au$ to become entrained by the wind (we explore radial variations using \crefalt{eq:width} to set $W$ in \cref{sec:radial_dependence}).

For grains that are not promptly returned to the disc, we integrate \cref{eq:fR,eq:fphi,eq:fz} to obtain the 3D trajectories in the wind. At every timestep the value of 
$\phi={\rm{tan}}^{-1}(z/\R)$ is used to map the dust grain onto the appropriate gas streamline passing
through this point. Pre-tabulated values of the self-similar gas streamline solution are used to identify $\phi$ with the distance along the streamline normalised to the base radius. Once we have the base radius and density normalisation we can then calculate the local gas density, streamline inclination, and poloidal gas 
velocity from the self-similar solution. The local azimuthal velocity of the gas is simply calculated using
conservation of specific angular momentum, assuming the gas is Keplerian at the flow base. These gas properties are then used to calculate the drag terms in the equations of motion.

The left panel in \cref{fig:dust_streamlines} depicts the projected $\R$-$z$ gas streamline and two example dust trajectories originating at $\R = 10\au$. In order to isolate the entrainment capabilities of the wind we assume the gas starts at the ionisation front with the ionised velocity and density without smoothing (although, for consistency with the discussion of dust delivery to the front, we will continue to label dust particles in terms of their smoothed Stokes number $\StIF$ defined in  \crefalt{eq:tildest}). 
The yellow dot-dashed curve represents the largest dust particle that can be launched without immediately re-intersecting the disc ($\smax = 6.85 \mum$). The Stokes number at the base of the yellow curve is $\StIF=\Stmax \sim 0.8$, slightly higher than our estimate based on \cref{eqn:force_ratio,eqn:chi_def}.
The initial path skims just above the disc surface (see the inset panel) before turning upward on a trajectory that is angled about halfway between the disc surface and the gas streamline. The orange solid line is for a dust particle that is ten times smaller (i.e. near the critical Stokes number $\Stcrit \sim 0.08$) and whose trajectory is more similar to the gas streamline. It then follows that the great majority of dust grains leaving the disc ($\StIF<\Stcrit$) follow trajectories that are close to that of the gas. The right panel shows the full 3D trajectories. The azimuthal drag provides a minor correction to the dust trajectory at each timestep that accumulates over time, but major differences do not appear until after the dust has already become unbound. This suggests that small deviations from our assumed gas flow may alter the detailed structure of the dust flow without compromising our overall conclusions.

We see that the dust grains of all sizes move monotonically away from the wind base. In particular, there are
no grain sizes for which grains begin to ascend and then re-descend to join the disc at larger radius. This
result is to be contrasted with what is found in the case of magneto-centrifugal winds by \citet{Giacalone/etal/2019} who find rain out of
grains even in the case of dust species with low Stokes number (where the short friction time assumption is valid).
The reason for this difference is likely to be the very different profiles of poloidal velocity (normalised to the local escape velocity) in the two cases. In  magnetocentrifugal winds, even tightly coupled grains would not achieve the escape velocity until they attained  a substantial
fraction of the Alfven radius, having then traversed  $\unsim 5$--$10 \times$ the radius of the flow base. Grains that are
somewhat less well coupled (Stokes number of order 1) cannot reach this point since they develop a downward terminal velocity. In the present case of an EUV-driven thermal wind, the gas launches more 
rapidly from the ionisation front and a tightly coupled grain would become unbound after traversing a distance 
of a few $10$s of $\%$ of the radius of the flow base. Even for more moderately coupled grains ($\Sti \sim 1$) for which the initial acceleration does not immediately return them to the disc,
the significantly more rapid flow close to the flow base results in them rapidly becoming unbound. For yet larger grains ($\Sti \in [1,\Stimax]$),
the net acceleration vector returns them promptly below the ionisation front; thus, these also do not contribute to radial transport of grains in the disc.

The above calculations assume that the dust is immediately exposed to the gas properties in the ionised wind (i.e. an infinitely thin ionisation front). If instead we start the dust at rest from $\zIF$ using the smoothed gas properties from our disc model, then the method of calculation has to be adjusted because the imposed structure of the front is not self-similar like the wind. To accommodate for this, at each timestep we smooth the appropriate gas streamline in the wind to its corresponding disc value at the base using the same general formalism as our disc model (e.g. \crefalt{eq:tildest}), only now the smoothing is done as a function of path length, the disc value is constant, and the wind value is variable. Depending on the assumed ionising luminosity $\Phii$ and front width $W$, we find that $\Stmax$ is reduced by $\unsim 20$--$40\%$ but that flow trajectories are otherwise little changed.

One type of trajectory that is not readily resolved in our calculations are those grains that are immediately returned to the disc. These grains are the generalisation of the hovering grains seen in 1D wind simulations \citepalias{Hutchison/etal/2016,Hutchison/Laibe/Maddison/2016}. It is plausible that once these grains re-enter the disc, the increased gas density will again propel them upwards, leading to an oscillation about the ionisation front. However, in contrast to the hovering grains in 1D, the non-zero radial velocity would lead to radial migration along the disc surface until they either become fully entrained in the wind or sink back down into the disc interior.
Although surface dust transport within the disc is potentially of interest \citep[as in the case of transport of  crystalline grains; e.g.][]{van-Boekel/etal/2005}, our conclusion that dust delivery shuts off at Stokes numbers somewhat below those for which dust entrainment in the wind is ineffective leads us to expect that this effect is not significant and we do not explore it further here.

\subsection{Dependence of dust wind properties on launch radius and ionising flux}
\label{sec:radial_dependence}

We have argued above that all of the dust delivered through the ionisation front at $R = 10\au$ can be entrained in the wind. Based on the fact that $\Stcrit$ (an indicative value for the Stokes number above which dust is not efficiently transported across the ionisation front) is an order of magnitude smaller than $\Stmax$ (the maximum Stokes number at the base of the ionised region for which grains are entrained in the flow), we further concluded that the the majority of dust in the wind follows trajectories that are similar to that of the gas. We now consider the radial scaling of $\Stmax$ and $\Stcrit$ to see whether these conclusions change with radius.

\begin{figure}
	\centering{\includegraphics[width=\columnwidth]{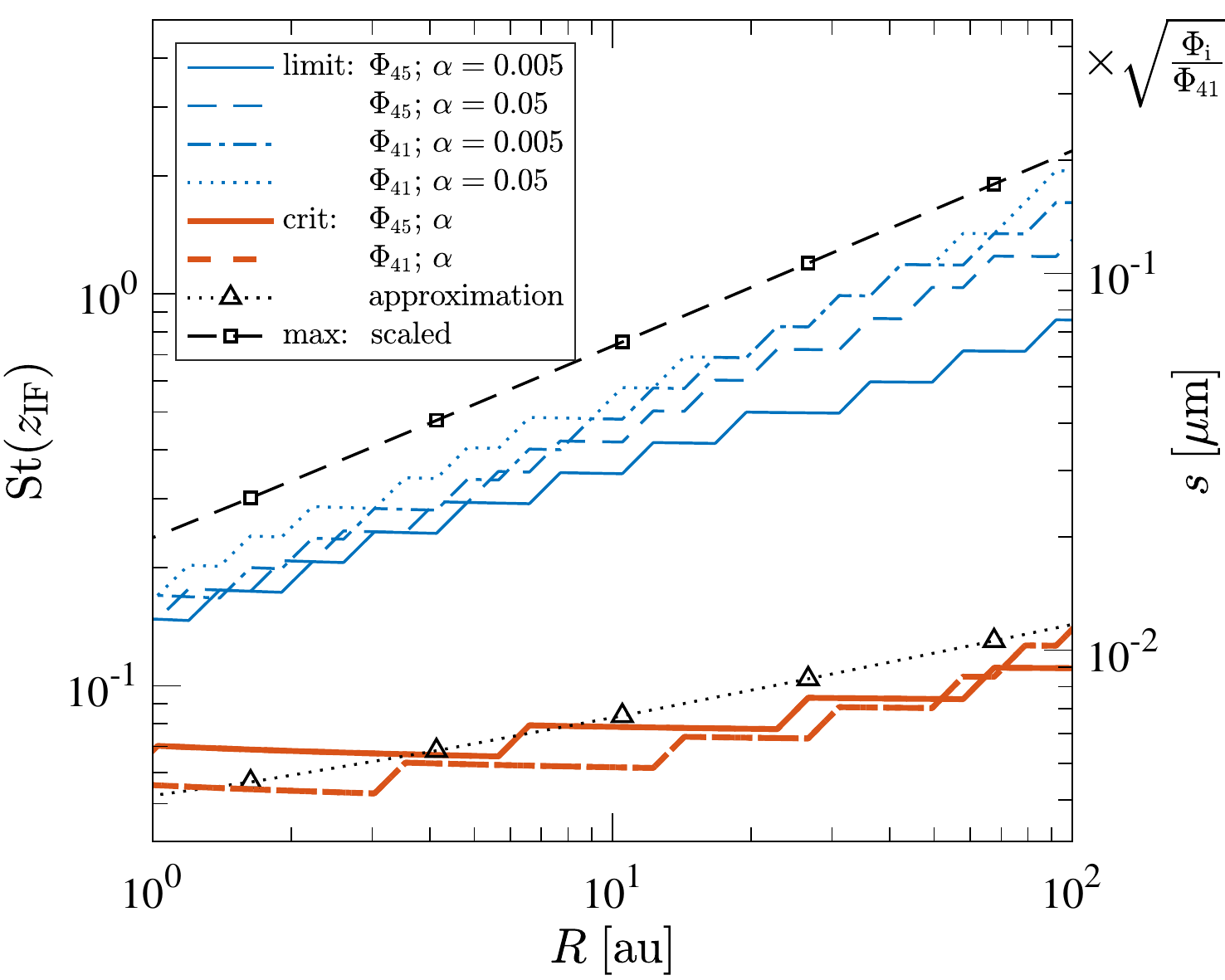}}
	\caption{Radial profiles for the three listed Stokes numbers (left axis) and associated grain sizes (right axis) when $\Phii = [10^{41},\,10^{45}] \ionflux$ and $\alpha = [0.005,\,0.05]$. The thick orange lines mark the {\it critical} Stokes number/grain size above which advection can no longer deliver grains to the ionisation front and is adequately approximated by \cref{eq:scrit} (black dotted line with triangular markers), despite the simplifying assumptions that go into its derivation. The thin blue lines represent the {\it upper limit} to the Stokes number/grain size that can be delivered to the wind by diffusion in the disc. Finally, the black dashed line with square markers is the theoretical {\it maximum entrainable} Stokes number/grain size based on properties in the wind, as derived from \cref{eqn:force_ratio,eqn:chi_def}.}
	\label{fig:Stmax_Stcrit}
\end{figure}

From \cref{eqn:force_ratio,eqn:chi_def}, we see that for a self-similar wind (where $\iIF$ and $\vgi$ are independent of radius), the entrainment criterion depends only on $\chi$ and hence on the product $\StIF \vK$. Since the upper limit to $\chi$ is constant, the maximum Stokes number at the ionisation front scales as $\Stmax \propto \sqrt{\R}$. Meanwhile, we can estimate $\Stcrit$ by first calculating $\scrit$ using the short friction time limit within the disc (i.e. setting $\vd(\zIF) = 0$ in \crefalt{eq:vdust_disc} and solving for $s$),
\begin{equation}
    \scrit = \vgn \frac{\cs \rhogn}{\rhoeff} \frac{\left( \R^2 + \zIF^2 \right)^{3/2}}{\G \Mstar \zIF} \propto \R^{0.25} M^{-0.5},
    \label{eq:scrit}
\end{equation}
and then using $\scrit$ to calculate the mid-point between the neutral and ionised Stokes numbers: $\StIF = 0.5(\tsn + \tsi)\OmegaK$. Conveniently, for the case that the density profile of the ionised wind base is given by \cref{eq:hollenbach_density} and where $\tsi \gg \tsn$, the radial scaling for $\Stcrit$ is equal to that of $\scrit$, allowing  us to plot the Stokes numbers and grain sizes together in the same figure.
We take advantage of this in \cref{fig:Stmax_Stcrit}, where we compare the radial scaling of $\Stcrit$ and $\scrit$ for different ionising luminosities $\Phii$ and turbulence levels $\alpha$. Note that provided $\tsi \gg \tsn$, $\Stcrit$ is independent of $\Phii$, an approximation that is well borne out in \cref{fig:Stmax_Stcrit}. We also plot $\Stlimit$, which is the maximum Stokes number at $\zIF$ for which our solutions in \cref{sec:dust_delivery} yield a non-zero flux at the ionisation front. In general we find that $\Stcrit < \Stlimit \lesssim \Stmax$ (the latter inequality being approximately equal if smoothing is factored into the maximum limit), supporting our earlier conclusion that the majority of grains leaving the disc follow similar trajectories to that of the gas for all radii relevant to photoevaporation.

We see from \cref{fig:Stmax_Stcrit} that there is a very weak dependence of $\Stlimit$ on $\Phii$ and $\alpha$. The flux dependence derives from its effect on the density of the wind base and hence the value of $W$, (\crefalt{eq:width}). 
As we saw previously in \cref{fig:f_ratio_plots,fig:F_St_limits}, the width $W$ affects $\Stlimit$ and the Stokes number corresponding to the peak dust flux (varying $\alpha$ produces similar effects). In the latter case, the peak flux can occur at Stokes numbers larger than $\Stcrit$ when $\Phii$ is small and $\R$ is large. As $\Stlimit$ approaches $\Stmax$, the deviation between dust and gas trajectories in the wind becomes more significant; however, the fact that this occurs at very low $\Phii$, where dust fluxes are in any case low, means that this is unlikely to be of observational consequence.{\footnote{Another parameter modulated by $\Phii$ is the height of the ionisation front $\zIF$, but the steep, near-Gaussian density gradient at four to five disc scale heights above the mid-plane renders this effect insignificant: since  $\rhogi \propto \sqrt{\Phii}$, a factor $100$ reduction in ionising luminosity only requires an order of magnitude change in $\rhogi$, which can be achieved by a very modest (order $10 \%$) increase in $\iIF$. Note that in order to maintain our assumption of constant $\iIF$, we have absorbed the variation with $\Phii$ into the disc sound speed and scale height, as detailed in \cref{sec:IF_location}.}} Since we have shown that both $\Stlimit$ and $\Stmax$ are very insensitive to $\alpha$ and $\Phii$, it follows that the corresponding grain sizes depend on $\Phii$ only via its influence on the base density of the ionised flow.
This implies that both $\smax$ and $\scrit$ scale with $\sqrt{\Phii}$, which explains the $\sqrt{\Phii/\Phi_{41}}$ dependency in the right-hand axis of \cref{fig:Stmax_Stcrit}. Finally, in real systems $\Phii$ is correlated with the stellar mass. While we do not vary $M$ in this study, we note there is an approximate $1/\sqrt{M}$ scaling in \cref{eq:scrit} that results from $\cs \approx \OmegaK \Hg$ and $\vgn \rhogn = {\rm constant}$.

\section{Implications for dusty photoevaporative winds}
\label{sec:implications}

We have shown that any grain that enters the ionised flow with a Stokes number less than a threshold value $\Stmax$ will be entrained in the flow. 
At $\R = 10\au$, this includes grains that are $\lesssim 7 \mum$ for $\Phi_{45}$ ($\lesssim 0.07 \mum$ for $\Phi_{41}$).
Above this threshold, dust re-enters the disc locally, although it may be possible for Stokes numbers very close to $\Stmax$ to skim along the disc surface until conditions are more favourable for entrainment (a process that we do not model here).
Grains that are less than around $10 \%$ of $\Stmax$ follow streamlines that are close to that of the gas, but all grains that are initially entrained in the wind rapidly achieve escape velocity and do not later rain out onto the disc at larger radii.
            
We have found that the diffusive description typically used to model the turbulent transport of dust in the disc interior leads to some non-intuitive results when applied to a near-discontinuous ionisation front. Namely, both the flux and size distribution of dust leaving the disc at a particular radius are sensitive to processes occurring on the length scale of the front. We presented a suite of simulations at $\R = 10 \au$ corresponding to different assumed front widths that show that the efficacy of mixing at the ionisation front is bracketed by two limits, the advection limit and the diffusive limit, as set out in \cref{sec:preliminaries}. 

In order to focus more on parameters relevant to real discs, in \cref{sec:radial_dependence} we switched to using \cref{eq:width} to approximate the physical width of the ionisation front. At $\R = 10\au$ we found $W \simeq 0.002 \au$ for $\Phi_{45}$ (appropriate for a very luminous Herbig Ae/Be star), which corresponds most closely to the purple lines in \cref{fig:F_St_limits}. Comparing the two turbulence cases reveals flux variations on the order of a few (except near the size $\slimit$ where the flux drops sharply to zero) and that
$\slimit$, which shifts from $\unsim2$ to $3 \mum$, is not strongly dependent on the level of turbulence. In both cases the dust flux is well below the scaled value of the gas flux even for grains sizes substantially less than 
$s \sim \scrit = 0.72 \mum$. Further comparison with \cref{fig:F_W_limits} suggests that, at this value of $W$, most grains are in the diffusive limit when $\alpha = 0.05$ and an intermediate state between the diffusive and advective limits for $\alpha = 0.005$. We emphasise that the suppression of the dust flux in these calculations is a consequence of the strong diffusion that
results from applying the advection-diffusion equation across a very narrow front (see discussion in \cref{sec:preliminaries}).

Of course not all stars encircled by protoplanetary discs have such large EUV luminosities. Lowering the ionising flux to $\Phi_{41}$ (appropriate for a low-mass T Tauri star) increases $W$ by two orders of magnitude (e.g. $W \simeq 0.15 \au$ at $\R = 10 \au$). Importantly, however, there is a commensurate shift in the width at which the crossover between the advective and diffusive limits takes place. Therefore, the general trends and conclusions for $\Phi_{45}$ and $\Phi_{41}$ are similar, although  there is some minor, order unity, variation in the radial profiles for $\Stlimit$ (see \cref{fig:Stmax_Stcrit}) and the Stokes number corresponding to the peak flux. In contrast, mapping these Stokes numbers onto grain sizes results in a $\sqrt{\Phii}$ dependence (see the right-hand scale of \cref{fig:Stmax_Stcrit}). It is also interesting to note that $W$ inherits its radial dependence from $\rhog(\zIF) \propto \R^{-1.5}$. It then follows that $W/\zIF \propto \sqrt{\R}$ and that the ionisation front in the inner disc is more prone to the limiting influences of diffusion than the outer disc.

One caveat to the arguments above is that \cref{eq:width} is calculated using the smoothed gas density, which is much closer to $\rhogi$ than $\rhogn$ (a natural consequence of mass conservation caused by the rapid acceleration of the gas across the ionisation front). However, since $\rhog(\zIF)$ and $\rhogn$ bracket nearly the entire drop in density across the ionisation front, it stands to reason that the correct `physical' interpretation is also bracketed by the effects predicted by each density. If we instead insert $\rhogn$ into \cref{eq:width}, we find that $W$ shrinks by almost two orders of magnitude (e.g. $W \simeq 2\times 10^{-5} \au$ at $\R = 10 \au$) and the diffusive effects at the ionisation front are magnified. Moreover, the radial dependence of $W/\zIF$ becomes a decreasing function of radius ($W/\zIF \propto \R^{-0.25}$), suggesting that diffusion becomes more important with $\R$. Clearly, $\rhogn$ predicts a more pessimistic limit on the dust flux and $\rhog(\zIF)$ the more optimistic, but both show evidence of $W$ regulating the dust flow through the ionisation front.

We caution once again that our results have been obtained using a simple prescription for diffusive mixing that is parameterised in terms of a turbulent mixing parameter $\alpha$. Furthermore, we have assumed that $\alpha$ transitions between its disc value to zero over a length scale $W$. Clearly the application of a diffusive description to a situation where the transport properties are changing over a length scale that is much smaller than any length scales associate with disc turbulence is questionable. It is for this reason that we de-emphasise a quantitative comparison. Nevertheless, whatever the appropriate microphysical description for dust mixing in the narrow region over which the gas is accelerated at the ionisation front, we argue that the diffusive limit would represent the appropriate limit if mixing prevented steep changes in the dust-to-gas ratio over the width of the front. As mentioned above, however, the attainment of this limit depends on the ability of dust grains to couple to the turbulent motions of the gas over the narrow front width, which is an imposed assumption in the formulation of \citet{Dubrulle/Morfill/Sterzik/1995}.

Given our ignorance of the details of microphysical mixing at the ionisation front, we will focus the rest of our discussion on the results that are more robust against these uncertainties.
\cref{fig:F_St_limits} illustrates the general point that $\StIF=\Stcrit$ (i.e. the Stokes number above which advection can no longer supply dust to the ionisation front) provides an order of magnitude indication of the Stokes number where the dust delivery turns down sharply, this limit being insensitive to the value of $W$.{\footnote{In our calculations the peak dust flux sometimes occurs slightly above $\Stcrit$ (e.g. at small $\Phii$ and large $\R$) but never by more than a factor of $\unsim 3$--$4$.}} 
This allows us to make general statements about the {\it range} of dust sizes that are delivered through the ionisation front. In contrast, the actual {\it value} of the dust flux (relative to the gas) is very sensitive to the details of mixing at the front. We therefore focus more on what we can say about the dust sizes that enter the wind.

The lower limit to the size distribution is relatively unimportant since we expect these particles to simply be advected away with the gas. The upper limit, on the other hand, can be defined in one of two ways. In an absolute sense, the upper limit at a particular launch radius is set by the competition between delivery and entrainment (i.e. ${\rm min}[\Stlimit,\Stmax]$). In practice, we find these limits are comparable -- particularly when the acceleration through the ionisation front is factored into the entrainment calculation. However, from \cref{fig:f_ratio_plots,fig:F_St_limits} we see that the dust flux is rapidly extinguished above $\Stcrit$, which leads us instead to define $\Stcrit$ as an effective maximum to our grain-size distribution in our wind. When combined with the trajectory calculations performed in \cref{fig:dust_streamlines}, we have reason to believe that the majority of grains delivered through the ionisation front are small enough to be well entrained in the wind. Pushing this simplification one step further to assume that the dust is stuck to the gas streamlines, then gradients in the grain-size distribution and associated optical properties of the wind purely derive from the variation in maximum grain size as a function of launch radius (roughly the grain size corresponding to $\StIF=\Stcrit$), as depicted in \cref{fig:Stmax_Stcrit}.

The common turnover in flux near $\Stcrit$ also allows us to place some constraints on the evolution of the dust-to-gas ratio in photoevaporating discs. The lack of dust loss for $\StIF > \Stcrit$  means that, per unit mass of gas lost in the wind, the minimum fraction of dust that remains in the disc is the fraction of dust in the mid-plane with $s > \scrit$; for a MRN size distribution spreading over many magnitudes in size, this fraction is given by $1-\sqrt{\scrit/\smaxmrn}$. Given that $\scrit$ is of (sub-)micron scale and that there is strong evidence from multi-wavelength sub-mm studies that grain growth has proceeded to mm sizes and above in protoplanetary discs \citep[e.g.][]{Testi/etal/2014}, it follows that a negligible fraction of dust mass is lost to the wind. While this seems to imply that photoevaporation could be a promising mechanism for driving up the dust-to-gas ratio in protoplanetary discs \citep[e.g.][]{Throop/Bally/2005},
it should be stressed that this conclusion only holds in regions where $\smaxmrn$ is large (e.g. has not been reduced by strong radial drift) and photoevaporation can remove a significant portion of the local gas mass (see \citealp{Sellek/Booth/Clarke/2020} for a similar conclusion in the case of dust entrained in winds driven by external FUV radiation in the outermost regions of protoplanetary discs). Enhancements in the dust-to-gas ratio by photoevaporation, if any, may therefore be limited to localised regions of the disc, such as dust traps, that can retain large dust grains over long timescales.

Having discussed the results from our own work, it is interesting to see how these results compare with previous studies of dusty EUV winds. At a radius of $\Rg = 9.46 \au$, \citet{Owen/Ercolano/Clarke/2011a} found a maximum grain size in the wind of around $2 \mum$.
This value is within $\unsim 3 \%$ of the maximum size implied by $\chi = \cot{\iIF}$ (the dashed line in \cref{fig:Stmax_Stcrit}), after correcting for the factor three difference in grain density and factor 100 difference in ionising luminosity ($\rhograin = 1 \gram \cm^{-3}$ and $\Phii = 10^{43} \ionflux$, respectively).
This agreement is unsurprising considering both calculations are based on the entrainment properties of the wind alone and we both source \citet{Hollenbach/etal/1994} for our ionisation front density. On the other hand, the curtailment of the dust flux between $\scrit$ and $\slimit$ is more analogous to the settling limit proposed by \citetalias{Hutchison/Laibe/Maddison/2016}, only here we have generalised the effect in two key ways. First, we account for the vertical drag from the flow feeding the wind and, secondly, we use a physical model to approximate the density, location, and finite extent of the ionisation front. Repeating the calculations of \citetalias{Hutchison/Laibe/Maddison/2016} using our disc parameters and ionisation front location, their model predicts a maximum grain size comparable to our $\scrit$ in the inner disc and about an order of magnitude smaller than our $\scrit$ in the outer disc (e.g. $\smax = [0.75, \;0.097] \mum$ at $\R=[1,\;100] \au$ assuming $\Phi_{45}$). These differences show the importance of accounting for the vertical gas flow in the disc and accurately modelling the ionisation front when obtaining the grain-size distribution in the wind. Ultimately, our calculations predict that the maximum grain size in photoevaporative winds is intermediate to the sizes proposed by \citet{Owen/Ercolano/Clarke/2011a} and \citetalias{Hutchison/Laibe/Maddison/2016}.

\subsection{Application to winds driven by non-ionising radiation}
\label{sec:non-ionising}

This paper focuses on winds driven by EUV radiation where the gas at the disc-wind interface is ionised and rapidly accelerated to near sonic speeds ($\unsim 0.44 \csi = 4.26 \kms$) over a spatially thin ionisation front. Using a diffusive model for turbulent diffusion, we have found that the extent to which small dust ($\StIF \lesssim 0.1$) can be advected with the wind is sensitive to the treatment of mixing at the ionisation front (see \cref{fig:F_St_limits}). We now consider what these calculations say about dust entrainment in other types of thermally driven photoevaporative winds. 

It is firstly important to stress that the radiation type `driving' the wind is that which provides the heating in the region where the gas transitions from subsonic to supersonic flow. Therefore, for example, a disc exposed to a mixture of EUV and FUV radiation can be driven, in this sense, by FUV radiation but also be mainly heated by EUV radiation at a point far out in the flow \citep{Johnstone/Hollenbach/Bally/1998}, a phenomenon that gives rise to the well-known offset ionisation fronts observed in the proplyds in the Orion Nebula Cluster and elsewhere \citep{ODell/Wen/Hu/1993,Kim/etal/2016}. In such cases, dust entrainment and acceleration to beyond the escape velocity is achieved at points in the flow that lie far interior to the ionisation front. Thus the effects observed in our simulations, where the ionisation front can throttle back the entrainment of dust, are least relevant to FUV driven winds. In contrast, winds that are driven by XEUV radiation still exhibit a spatially sharp transition at the disc-wind interface \citep{Picogna/etal/2019}, potentially sharp enough to experience some of the effects observed in this study. 

For winds that lack a sharply defined front, we can expect the gas temperature to vary smoothly along the flow trajectory \citep{Facchini/Clarke/Bisbas/2016,Owen/Clarke/Ercolano/2012}. 
The resulting dust entrainment may therefore be similar to our solutions with large $W$ where the gas is accelerated gradually over a substantial fraction of the total disc height. The larger densities and faster velocities in the acceleration region lead to an increase in both $\Stcrit$ and the flux, the latter being close to the advective limit nearly up to $\Stcrit$ before being rapidly quenched. The combination of having a larger grain-size distribution at higher fluxes may indicate that, given the same gas flux, FUV winds have the potential to be more dust rich than winds that exhibit a sharp disc-wind interface.

The X-ray case has been investigated by \citet{Franz/etal/2020} for a mass-loss rate at $\R=10\au$ that, in our model, corresponds to an EUV luminosity of $\unsim 10^{43} \ionflux$. The focus of their study is on the entrainment properties of the wind, which they obtain by calculating trajectories of dust particles introduced at the base of the heated flow. Since trajectories alone are insufficient to make statements regarding flux, their results are best compared with $\smax$ in our model.
Accounting for the different parameters employed by \citet{Franz/etal/2020} (stellar mass, gas density and velocity, sound speed, local inclination angle of the disc surface, and intrinsic grain density), setting $\chi = \cot{\iIF}$ gives a maximum size of $\smax \sim 7.6 \mum$ at $\R = 10 \au$ -- only a little smaller than the $9\mum$ they report at the same radius. The approximate agreement of $\smax$ to XEUV calculations can be attributed to the fact that (i) XEUV winds have a sharp disc-wind interface with clearly defined wind properties{\footnote{It would be more difficult to apply $\smax$ to FUV winds where the acceleration of the wind is more gradual.}} at the base of the flow and (ii) $\smax$ is a dynamical limit, derived without reference to the heating mechanism generating the wind. As a final point of interest, \citet{Franz/etal/2020} observed some of their larger grains reconnecting with the disc at large radii ($\R > 100 \au$), following a rapid decline in the base density of the wind. It is likely we do not observe such trajectories in our wind model because of the fixed power-law relation we assume for $\rhogi$ in \cref{eq:hollenbach_density}. Rainout of grains in photoevaporative winds may therefore only occur in regions where the radial base density profile is very steep.

\section{Conclusions}
\label{sec:conclusions}

In this study we have coupled a turbulent gas disc with an inclined self-similar EUV-driven photoevaporative wind and attempted to track the flow of dust as it travels from the disc mid-plane into the wind. By solving the fluid equations for the dust in the disc and the equations of motion for the dust in the wind, we are able to explore aspects of both {\it delivery} and {\it entrainment} of dust in EUV winds. \citetalias{Hutchison/Laibe/Maddison/2016} previously argued that delivery of dust to the ionisation front (in their case via diffusion in a static disc) controlled the upper size limit of dust grains in photoevaporative winds for a large fraction of the disc. In the present study, we find that including the advection of dust within the gas flow feeding the wind helps to improve the diffusive delivery of large grains to the ionisation front, bringing the maximum deliverable size from the disc and maximum entrainable size from the wind nearly into agreement. However, while the deliverable size limit is increased, the exiting flux of grains near this limit remains small. Our simulations instead point to an effective maximum set by the steep turnover in dust flux near what we call the critical grain size, which is intermediate to the sizes proposed earlier in the literature for EUV winds \citep[][\citetalias{Hutchison/Laibe/Maddison/2016}]{Owen/Ercolano/Clarke/2011a}. Importantly, this turnover at the critical size limit holds over a wide range of ionising luminosities, turbulence strengths, and ionisation front widths.

The critical grain size corresponds to the size at which the advective flow of dust is first interrupted by the dust velocity reversing directions (usually near the ionisation front).
For the fiducial case modeled here (a solar-mass star with an EUV luminosity of $10^{45} \ionflux$, corresponding to a very luminous Herbig Ae/Be star, and an intrinsic grain density of $3 \gram \cm^{-3}$), the critical grain size is $\lesssim 1 \mum$. This is sufficiently small compared to typical range of grain sizes that are present in the disc mid-plane such that only a very small fraction of the dust mass is lost in the wind. Thus EUV driven photoevaporation provides a mechanism for driving up the dust-to-gas ratio in discs (for a discussion on how our results are likely to affect estimates of dust transport in photoevaporative winds driven by other types of energetic radiation, see \cref{sec:non-ionising}). Alternatively, for EUV luminosities typical of T Tauri stars ($\unsim 10^{41} \ionflux$) the critical grain size delivered to the ionisation front is $\lesssim 10^{-2}\mum$, implying that EUV winds generated by T Tauri stars should be essentially {\it dust free}. Therefore, any observations of dust in halos around T Tauri stars would likely be due to alternative mechanisms (e.g. magneto-centrifugal winds or infall).
    
To summarise other key results:
\begin{enumerate}
\setlength\itemsep{0em}
    \item We find that essentially all grain sizes delivered to the ionisation front are entrained in the wind and eventually escape to infinity.
    \item The maximum entrainable grain size in EUV winds is set by the wind properties at the surface of the disc rather than at some later point in the wind. Thus, in contrast to hydro-magnetic winds \citep{Giacalone/etal/2019}, we do not find evidence for fall-back of grains onto the disc at larger radii for our assumed wind-base profile. However, it may be possible for steeper base density profiles, particularly in regions where photoevaporation experiences a sudden drop in efficiency \citep[as seen in][]{Franz/etal/2020}.
    \item Taking advantage of the fact that the maximum size limit is set at the surface of the disc, we have derived a simple entrainment criterion based on whether the ratio of the vertical to radial forces result in the dust immediately intersecting with the inclined surface of the disc ($\chi > \cot{\iIF}$; see \cref{sec:wind_solution}). Importantly, this criterion is derived without reference to the heating mechanism in the wind and we find that it is in good agreement with all previous predictions in the literature.
    \item Large grains near the maximum limit do not vertically sink back into the disc but re-enter with an outward radial velocity, potentially allowing them to skim radially along the surface of the neutral wind until they are either entrained or become permanently trapped by the disc.
    \item The drag from the vertical flow created by photoevaporation aids in lofting grains into the surface layers of the disc. Even if these grains are not ultimately lost to the wind, they could enhance  mechanisms such as the one recently proposed by \citet{Owen/Kollmeier/2019} regarding the removal of surface grains by radiation pressure in transition discs.
    \item For grains whose velocity remains positive throughout the flow (i.e. are always delivered to the ionisation front), we find that the flux is sensitive to the efficiency of diffusive mixing in the vicinity of the ionisation front. We have identified two analytical limits that explain the observed flux variation in our model:
        \begin{itemize}
        \setlength\itemsep{0.5em}
            \item Diffusive limit: characterised by strong diffusive mixing that leads to low dust delivery rates to the ionisation front. Although more prominent for large grains, small grains are also susceptible.
            \item Advective limit: characterised by weak diffusive mixing, allowing advection to deliver grains to the ionisation front in the ratio in which they are present in the disc mid-plane.
        \end{itemize}
    In deriving these limits, we have uncovered a subtle inconsistency in using standard disc turbulence models (where the relevant timescale is the local dynamical time of the disc) to model the diffusivity of dust that passes through a spatially thin ionisation front on much shorter timescales.
    Further modeling of the microphysics at the ionisation front is required to determine which limit is more physically plausible. However, since the fraction of the mid-plane dust that is delivered to the ionisation front is low even in the advection limit (the maximum attainable flux), this uncertainty does not affect the conclusions drawn above.
\end{enumerate}

\section*{Acknowledgments}

This work has been carried out within the framework of the National Centre for Competence in Research PlanetS, supported by the Swiss National Science Foundation. We thank Gavin Coleman, Rainer Schr{\"a}pler, Richard
Booth, Sebastien Krijt and James Sutherland for fruitful discussions. We also thank the anonymous referee whose insightful comments helped improve this paper. MAH acknowledges the financial support of the SNSF. CJC is grateful for funding from  the European Union’s Horizon 2020 research and innovation programme under the Marie Sklodowska-Curie grant agreement No 823823 (DUSTBUSTERS) and for support from the STFC consolidated grant ST/S000623/1. CJC also
gratefully acknowledges hospitality from LMU, funded by the CAM-LMU initiative, during completion of this project.

\section*{Data availability}

The data underlying this article will be shared on reasonable request to the corresponding author.



\bibliographystyle{mnras}
\bibliography{bibliography}

\begin{thebibliography}{}
\makeatletter
\relax
\def\mn@urlcharsother{\let\do\@makeother \do\$\do\&\do\#\do\^\do\_\do\%\do\~}
\def\mn@doi{\begingroup\mn@urlcharsother \@ifnextchar [ {\mn@doi@}
  {\mn@doi@[]}}
\def\mn@doi@[#1]#2{\def\@tempa{#1}\ifx\@tempa\@empty \href
  {http://dx.doi.org/#2} {doi:#2}\else \href {http://dx.doi.org/#2} {#1}\fi
  \endgroup}
\def\mn@eprint#1#2{\mn@eprint@#1:#2::\@nil}
\def\mn@eprint@arXiv#1{\href {http://arxiv.org/abs/#1} {{\tt arXiv:#1}}}
\def\mn@eprint@dblp#1{\href {http://dblp.uni-trier.de/rec/bibtex/#1.xml}
  {dblp:#1}}
\def\mn@eprint@#1:#2:#3:#4\@nil{\def\@tempa {#1}\def\@tempb {#2}\def\@tempc
  {#3}\ifx \@tempc \@empty \let \@tempc \@tempb \let \@tempb \@tempa \fi \ifx
  \@tempb \@empty \def\@tempb {arXiv}\fi \@ifundefined
  {mn@eprint@\@tempb}{\@tempb:\@tempc}{\expandafter \expandafter \csname
  mn@eprint@\@tempb\endcsname \expandafter{\@tempc}}}

\bibitem[\protect\citeauthoryear{{Adams}, {Hollenbach}, {Laughlin}  \&
  {Gorti}}{{Adams} et~al.}{2004}]{Adams/etal/2004}
{Adams} F.~C.,  {Hollenbach} D.,  {Laughlin} G.,   {Gorti} U.,  2004, \apj,
  611, 360

\bibitem[\protect\citeauthoryear{{Alexander} \& {Pascucci}}{{Alexander} \&
  {Pascucci}}{2012}]{Alexander/Pascucci/2012}
{Alexander} R.~D.,  {Pascucci} I.,  2012, \mnras, 422, L82

\bibitem[\protect\citeauthoryear{{Bai}, {Ye}, {Goodman}  \& {Yuan}}{{Bai}
  et~al.}{2016}]{Bai/etal/2016}
{Bai} X.-N.,  {Ye} J.,  {Goodman} J.,   {Yuan} F.,  2016, \apj, 818, 152

\bibitem[\protect\citeauthoryear{{Birnstiel}, {Fang}  \&
  {Johansen}}{{Birnstiel} et~al.}{2016}]{Birnstiel/Fang/Johansen/2016}
{Birnstiel} T.,  {Fang} M.,   {Johansen} A.,  2016, \ssr

\bibitem[\protect\citeauthoryear{{Bisbas}, {Bell}, {Viti}, {Yates}  \&
  {Barlow}}{{Bisbas} et~al.}{2012}]{Bisbas/etal/2012}
{Bisbas} T.~G.,  {Bell} T.~A.,  {Viti} S.,  {Yates} J.,   {Barlow} M.~J.,
  2012, \mnras, 427, 2100

\bibitem[\protect\citeauthoryear{{Boss}}{{Boss}}{1997}]{Boss/1997}
{Boss} A.~P.,  1997, Science, 276, 1836

\bibitem[\protect\citeauthoryear{{Calvet} et~al.,}{{Calvet}
  et~al.}{2005}]{Calvet/etal/2005}
{Calvet} N.,  et~al., 2005, \apjl, 630, L185

\bibitem[\protect\citeauthoryear{{Carrera}, {Gorti}, {Johansen}  \&
  {Davies}}{{Carrera} et~al.}{2017}]{Carrera/etal/2017}
{Carrera} D.,  {Gorti} U.,  {Johansen} A.,   {Davies} M.~B.,  2017, \apj, 839,
  16

\bibitem[\protect\citeauthoryear{{Charnoz}, {Fouchet}, {Aleon}  \&
  {Moreira}}{{Charnoz} et~al.}{2011}]{Charnoz/etal/2011}
{Charnoz} S.,  {Fouchet} L.,  {Aleon} J.,   {Moreira} M.,  2011, \apj, 737, 33

\bibitem[\protect\citeauthoryear{{Chiang} \& {Murray-Clay}}{{Chiang} \&
  {Murray-Clay}}{2007}]{Chiang/Murray-Clay/2007}
{Chiang} E.,  {Murray-Clay} R.,  2007, Nature Physics, 3, 604

\bibitem[\protect\citeauthoryear{{Clarke} \& {Alexander}}{{Clarke} \&
  {Alexander}}{2016}]{Clarke/Alexander/2016}
{Clarke} C.~J.,  {Alexander} R.~D.,  2016, \mnras, 460, 3044

\bibitem[\protect\citeauthoryear{{Clarke}, {Gendrin}  \& {Sotomayor}}{{Clarke}
  et~al.}{2001}]{Clarke/Gendrin/Sotomayor/2001}
{Clarke} C.~J.,  {Gendrin} A.,   {Sotomayor} M.,  2001, \mnras, 328, 485

\bibitem[\protect\citeauthoryear{{Dubrulle}, {Morfill}  \&
  {Sterzik}}{{Dubrulle} et~al.}{1995}]{Dubrulle/Morfill/Sterzik/1995}
{Dubrulle} B.,  {Morfill} G.,   {Sterzik} M.,  1995, \icarus, 114, 237

\bibitem[\protect\citeauthoryear{{Dullemond} \& {Dominik}}{{Dullemond} \&
  {Dominik}}{2005}]{Dullemond/Dominik/2005}
{Dullemond} C.~P.,  {Dominik} C.,  2005, \aap, 434, 971

\bibitem[\protect\citeauthoryear{{Epstein}}{{Epstein}}{1924}]{Epstein/1924}
{Epstein} P.~S.,  1924, Physical Review, 23, 710

\bibitem[\protect\citeauthoryear{{Ercolano} \& {Pascucci}}{{Ercolano} \&
  {Pascucci}}{2017}]{Ercolano/Pascucci/2017}
{Ercolano} B.,  {Pascucci} I.,  2017, \mn@doi [Royal Society Open Science]
  {10.1098/rsos.170114}, 4, 170114

\bibitem[\protect\citeauthoryear{{Ercolano} \& {Rosotti}}{{Ercolano} \&
  {Rosotti}}{2015}]{Ercolano/Rosotti/2015}
{Ercolano} B.,  {Rosotti} G.,  2015, \mnras, 450, 3008

\bibitem[\protect\citeauthoryear{{Ercolano}, {Drake}, {Raymond}  \&
  {Clarke}}{{Ercolano} et~al.}{2008}]{Ercolano/etal/2008}
{Ercolano} B.,  {Drake} J.~J.,  {Raymond} J.~C.,   {Clarke} C.~C.,  2008, \apj,
  688, 398

\bibitem[\protect\citeauthoryear{{Ercolano}, {Koepferl}, {Owen}  \&
  {Robitaille}}{{Ercolano} et~al.}{2015}]{Ercolano/etal/2015}
{Ercolano} B.,  {Koepferl} C.,  {Owen} J.,   {Robitaille} T.,  2015, \mnras,
  452, 3689

\bibitem[\protect\citeauthoryear{{Facchini}, {Clarke}  \& {Bisbas}}{{Facchini}
  et~al.}{2016}]{Facchini/Clarke/Bisbas/2016}
{Facchini} S.,  {Clarke} C.~J.,   {Bisbas} T.~G.,  2016, \mnras, 457, 3593

\bibitem[\protect\citeauthoryear{{Franz}, {Picogna}, {Ercolano}  \&
  {Birnstiel}}{{Franz} et~al.}{2020}]{Franz/etal/2020}
{Franz} R.,  {Picogna} G.,  {Ercolano} B.,   {Birnstiel} T.,  2020, \mn@doi
  [\aap] {10.1051/0004-6361/201936615}, 635, A53

\bibitem[\protect\citeauthoryear{{Fromang} \& {Nelson}}{{Fromang} \&
  {Nelson}}{2009}]{Fromang/Nelson/2009}
{Fromang} S.,  {Nelson} R.~P.,  2009, \aap, 496, 597

\bibitem[\protect\citeauthoryear{{Fromang} \& {Papaloizou}}{{Fromang} \&
  {Papaloizou}}{2006}]{Fromang/Papaloizou/2006}
{Fromang} S.,  {Papaloizou} J.,  2006, \aap, 452, 751

\bibitem[\protect\citeauthoryear{{Giacalone}, {Teitler}, {K{\"o}nigl}, {Krijt}
  \& {Ciesla}}{{Giacalone} et~al.}{2019}]{Giacalone/etal/2019}
{Giacalone} S.,  {Teitler} S.,  {K{\"o}nigl} A.,  {Krijt} S.,   {Ciesla} F.~J.,
   2019, \mn@doi [\apj] {10.3847/1538-4357/ab311a}, 882, 33

\bibitem[\protect\citeauthoryear{{Goldreich} \& {Tremaine}}{{Goldreich} \&
  {Tremaine}}{1980}]{Goldreich/Tremaine/1980}
{Goldreich} P.,  {Tremaine} S.,  1980, \apj, 241, 425

\bibitem[\protect\citeauthoryear{{Gorti}, {Dullemond}  \& {Hollenbach}}{{Gorti}
  et~al.}{2009}]{Gorti/Dullemond/Hollenbach/2009}
{Gorti} U.,  {Dullemond} C.~P.,   {Hollenbach} D.,  2009, \apj, 705, 1237

\bibitem[\protect\citeauthoryear{{Haisch}, {Lada}  \& {Lada}}{{Haisch}
  et~al.}{2001}]{Haisch/Lada/Lada/2001}
{Haisch} Jr. K.~E.,  {Lada} E.~A.,   {Lada} C.~J.,  2001, \apjl, 553, L153

\bibitem[\protect\citeauthoryear{{Hern{\'a}ndez} et~al.,}{{Hern{\'a}ndez}
  et~al.}{2007}]{Hernandez/etal/2007}
{Hern{\'a}ndez} J.,  et~al., 2007, \apj, 671, 1784

\bibitem[\protect\citeauthoryear{{Hollenbach}, {Johnstone}, {Lizano}  \&
  {Shu}}{{Hollenbach} et~al.}{1994}]{Hollenbach/etal/1994}
{Hollenbach} D.,  {Johnstone} D.,  {Lizano} S.,   {Shu} F.,  1994, \apj, 428,
  654

\bibitem[\protect\citeauthoryear{{Hutchison} \& {Laibe}}{{Hutchison} \&
  {Laibe}}{2016}]{Hutchison/Laibe/2016}
{Hutchison} M.~A.,  {Laibe} G.,  2016, \pasa, 33, e014

\bibitem[\protect\citeauthoryear{{Hutchison}, {Price}, {Laibe}  \&
  {Maddison}}{{Hutchison} et~al.}{2016a}]{Hutchison/etal/2016}
{Hutchison} M.~A.,  {Price} D.~J.,  {Laibe} G.,   {Maddison} S.~T.,  2016a,
  \mnras, 461, 742

\bibitem[\protect\citeauthoryear{{Hutchison}, {Laibe}  \&
  {Maddison}}{{Hutchison} et~al.}{2016b}]{Hutchison/Laibe/Maddison/2016}
{Hutchison} M.~A.,  {Laibe} G.,   {Maddison} S.~T.,  2016b, \mnras, 463, 2725

\bibitem[\protect\citeauthoryear{{Hutchison}, {Price}  \& {Laibe}}{{Hutchison}
  et~al.}{2018}]{Hutchison/Price/Laibe/2018}
{Hutchison} M.,  {Price} D.~J.,   {Laibe} G.,  2018, \mnras, 476, 2186

\bibitem[\protect\citeauthoryear{{Ikoma}, {Nakazawa}  \& {Emori}}{{Ikoma}
  et~al.}{2000}]{Ikoma/Nakazawa/Emori/2000}
{Ikoma} M.,  {Nakazawa} K.,   {Emori} H.,  2000, \apj, 537, 1013

\bibitem[\protect\citeauthoryear{{Inutsuka}, {Machida}  \&
  {Matsumoto}}{{Inutsuka} et~al.}{2010}]{Inutsuka/Machida/Matsumoto/2010}
{Inutsuka} S.-i.,  {Machida} M.~N.,   {Matsumoto} T.,  2010, \apjl, 718, L58

\bibitem[\protect\citeauthoryear{{Jennings}, {Ercolano}  \&
  {Rosotti}}{{Jennings} et~al.}{2018}]{Jennings/Ercolano/Rosotti/2018}
{Jennings} J.,  {Ercolano} B.,   {Rosotti} G.~P.,  2018, \mn@doi [\mnras]
  {10.1093/mnras/sty964}, 477, 4131

\bibitem[\protect\citeauthoryear{{Johansen} \& {Klahr}}{{Johansen} \&
  {Klahr}}{2005}]{Johansen/Klahr/2005}
{Johansen} A.,  {Klahr} H.,  2005, \apj, 634, 1353

\bibitem[\protect\citeauthoryear{{Johansen} \& {Youdin}}{{Johansen} \&
  {Youdin}}{2007}]{Johansen/Youdin/2007}
{Johansen} A.,  {Youdin} A.,  2007, \apj, 662, 627

\bibitem[\protect\citeauthoryear{{Johnstone}, {Hollenbach}  \&
  {Bally}}{{Johnstone} et~al.}{1998}]{Johnstone/Hollenbach/Bally/1998}
{Johnstone} D.,  {Hollenbach} D.,   {Bally} J.,  1998, \apj, 499, 758

\bibitem[\protect\citeauthoryear{{Kim}, {Clarke}, {Fang}  \& {Facchini}}{{Kim}
  et~al.}{2016}]{Kim/etal/2016}
{Kim} J.~S.,  {Clarke} C.~J.,  {Fang} M.,   {Facchini} S.,  2016, \mn@doi
  [\apjl] {10.3847/2041-8205/826/1/L15}, 826, L15

\bibitem[\protect\citeauthoryear{{Krauss}, {Wurm}, {Mousis}, {Petit}, {Horner}
  \& {Alibert}}{{Krauss} et~al.}{2007}]{Krauss/etal/2007}
{Krauss} O.,  {Wurm} G.,  {Mousis} O.,  {Petit} J.-M.,  {Horner} J.,
  {Alibert} Y.,  2007, \aap, 462, 977

\bibitem[\protect\citeauthoryear{{Kuiper}}{{Kuiper}}{1951}]{Kuiper/1951}
{Kuiper} G.~P.,  1951, in {Hynek} J.~A.,  ed., 50th Anniversary of the Yerkes
  Observatory and Half a Century of Progress in Astrophysics. p.~357

\bibitem[\protect\citeauthoryear{{Laibe}}{{Laibe}}{2014}]{Laibe/2014}
{Laibe} G.,  2014, \mnras, 437, 3037

\bibitem[\protect\citeauthoryear{{Laibe}, {Gonzalez}  \& {Maddison}}{{Laibe}
  et~al.}{2012}]{Laibe/Gonzalez/Maddison/2012}
{Laibe} G.,  {Gonzalez} J.-F.,   {Maddison} S.~T.,  2012, \aap, 537, A61

\bibitem[\protect\citeauthoryear{{Lynden-Bell} \& {Pringle}}{{Lynden-Bell} \&
  {Pringle}}{1974}]{Lynden-Bell/Pringle/1974}
{Lynden-Bell} D.,  {Pringle} J.~E.,  1974, \mnras, 168, 603

\bibitem[\protect\citeauthoryear{{Mamajek}}{{Mamajek}}{2009}]{Mamajek/2009}
{Mamajek} E.~E.,  2009, in {Usuda} T.,  {Tamura} M.,   {Ishii} M.,  eds,
  American Institute of Physics Conference Series Vol. 1158, American Institute
  of Physics Conference Series. pp 3--10

\bibitem[\protect\citeauthoryear{{Marsh} \& {Mahoney}}{{Marsh} \&
  {Mahoney}}{1992}]{Marsh/Mahoney/1992}
{Marsh} K.~A.,  {Mahoney} M.~J.,  1992, \apjl, 395, L115

\bibitem[\protect\citeauthoryear{{Mathis}, {Rumpl}  \& {Nordsieck}}{{Mathis}
  et~al.}{1977}]{Mathis/Rumpl/Nordsieck/1977}
{Mathis} J.~S.,  {Rumpl} W.,   {Nordsieck} K.~H.,  1977, \apj, 217, 425

\bibitem[\protect\citeauthoryear{{Matsuyama}, {Johnstone}  \&
  {Hartmann}}{{Matsuyama} et~al.}{2003}]{Matsuyama/Johnstone/Hartmann/2003}
{Matsuyama} I.,  {Johnstone} D.,   {Hartmann} L.,  2003, \apj, 582, 893

\bibitem[\protect\citeauthoryear{{Mordasini}, {Alibert}, {Georgy}, {Dittkrist},
  {Klahr}  \& {Henning}}{{Mordasini} et~al.}{2012}]{Mordasini/etal/2012}
{Mordasini} C.,  {Alibert} Y.,  {Georgy} C.,  {Dittkrist} K.-M.,  {Klahr} H.,
  {Henning} T.,  2012, \aap, 547, A112

\bibitem[\protect\citeauthoryear{{Nayakshin}}{{Nayakshin}}{2010}]{Nayakshin/2010}
{Nayakshin} S.,  2010, \mnras, 408, L36

\bibitem[\protect\citeauthoryear{{O'dell}, {Wen}  \& {Hu}}{{O'dell}
  et~al.}{1993}]{ODell/Wen/Hu/1993}
{O'dell} C.~R.,  {Wen} Z.,   {Hu} X.,  1993, \apj, 410, 696

\bibitem[\protect\citeauthoryear{{Owen} \& {Kollmeier}}{{Owen} \&
  {Kollmeier}}{2019}]{Owen/Kollmeier/2019}
{Owen} J.~E.,  {Kollmeier} J.~A.,  2019, \mn@doi [\mnras]
  {10.1093/mnras/stz1591}, \href
  {https://ui.adsabs.harvard.edu/abs/2019MNRAS.487.3702O} {487, 3702}

\bibitem[\protect\citeauthoryear{{Owen}, {Ercolano}  \& {Clarke}}{{Owen}
  et~al.}{2011a}]{Owen/Ercolano/Clarke/2011a}
{Owen} J.~E.,  {Ercolano} B.,   {Clarke} C.~J.,  2011a, \mnras, 411, 1104

\bibitem[\protect\citeauthoryear{{Owen}, {Ercolano}  \& {Clarke}}{{Owen}
  et~al.}{2011b}]{Owen/Ercolano/Clarke/2011b}
{Owen} J.~E.,  {Ercolano} B.,   {Clarke} C.~J.,  2011b, \mnras, 412, 13

\bibitem[\protect\citeauthoryear{{Owen}, {Clarke}  \& {Ercolano}}{{Owen}
  et~al.}{2012}]{Owen/Clarke/Ercolano/2012}
{Owen} J.~E.,  {Clarke} C.~J.,   {Ercolano} B.,  2012, \mnras, 422, 1880

\bibitem[\protect\citeauthoryear{{Picogna}, {Ercolano}, {Owen}  \&
  {Weber}}{{Picogna} et~al.}{2019}]{Picogna/etal/2019}
{Picogna} G.,  {Ercolano} B.,  {Owen} J.~E.,   {Weber} M.~L.,  2019, \mn@doi
  [\mnras] {10.1093/mnras/stz1166}, 487, 691

\bibitem[\protect\citeauthoryear{{Pollack}, {Hubickyj}, {Bodenheimer},
  {Lissauer}, {Podolak}  \& {Greenzweig}}{{Pollack}
  et~al.}{1996}]{Pollack/etal/1996}
{Pollack} J.~B.,  {Hubickyj} O.,  {Bodenheimer} P.,  {Lissauer} J.~J.,
  {Podolak} M.,   {Greenzweig} Y.,  1996, \icarus, 124, 62

\bibitem[\protect\citeauthoryear{{Quintana}, {Adams}, {Lissauer}  \&
  {Chambers}}{{Quintana} et~al.}{2007}]{Quintana/etal/2007}
{Quintana} E.~V.,  {Adams} F.~C.,  {Lissauer} J.~J.,   {Chambers} J.~E.,  2007,
  \apj, 660, 807

\bibitem[\protect\citeauthoryear{{Riols} \& {Lesur}}{{Riols} \&
  {Lesur}}{2018}]{Riols/Lesur/2018}
{Riols} A.,  {Lesur} G.,  2018, \aap, 617, A117

\bibitem[\protect\citeauthoryear{{Safronov}}{{Safronov}}{1969}]{Safronov/1969}
{Safronov} V.~S.,  1969, {Evolution of the protoplanetary cloud and formation
  of the earth and planets}.
Keter Publishing House, Jerusalem, p. 212

\bibitem[\protect\citeauthoryear{{Schr{\"a}pler} \& {Henning}}{{Schr{\"a}pler}
  \& {Henning}}{2004}]{Schrapler/Henning/2004}
{Schr{\"a}pler} R.,  {Henning} T.,  2004, \apj, 614, 960

\bibitem[\protect\citeauthoryear{{Sellek}, {Booth}  \& {Clarke}}{{Sellek}
  et~al.}{2020}]{Sellek/Booth/Clarke/2020}
{Sellek} A.~D.,  {Booth} R.~A.,   {Clarke} C.~J.,  2020, \mn@doi [\mnras]
  {10.1093/mnras/stz3528}, 492, 1279

\bibitem[\protect\citeauthoryear{{Shakura} \& {Sunyaev}}{{Shakura} \&
  {Sunyaev}}{1973}]{Shakura/Sunyaev/1973}
{Shakura} N.~I.,  {Sunyaev} R.~A.,  1973, \aap, 24, 337

\bibitem[\protect\citeauthoryear{Shampine \& Reichelt}{Shampine \&
  Reichelt}{1997}]{Shampine/Reichelt/1997}
Shampine L.~F.,  Reichelt M.~W.,  1997, SIAM Journal on Scientific Computing,
  18, 1

\bibitem[\protect\citeauthoryear{Shampine, Reichelt  \& Kierzenka}{Shampine
  et~al.}{1999}]{Shampine/Reichelt/Kierzenka/1999}
Shampine L.~F.,  Reichelt M.~W.,   Kierzenka J.~A.,  1999, SIAM Review, 41, 538

\bibitem[\protect\citeauthoryear{{Shi} \& {Chiang}}{{Shi} \&
  {Chiang}}{2014}]{Shi/Chiang/2014}
{Shi} J.-M.,  {Chiang} E.,  2014, \apj, 789, 34

\bibitem[\protect\citeauthoryear{{Suzuki} \& {Inutsuka}}{{Suzuki} \&
  {Inutsuka}}{2009}]{Suzuki/Inutsuka/2009}
{Suzuki} T.~K.,  {Inutsuka} S.-i.,  2009, \apjl, 691, L49

\bibitem[\protect\citeauthoryear{{Takeuchi} \& {Lin}}{{Takeuchi} \&
  {Lin}}{2002}]{Takeuchi/Lin/2002}
{Takeuchi} T.,  {Lin} D.~N.~C.,  2002, \apj, 581, 1344

\bibitem[\protect\citeauthoryear{{Takeuchi}, {Clarke}  \& {Lin}}{{Takeuchi}
  et~al.}{2005}]{Takeuchi/Clarke/Lin/2005}
{Takeuchi} T.,  {Clarke} C.~J.,   {Lin} D.~N.~C.,  2005, \apj, \href
  {http://adsabs.harvard.edu/abs/2005ApJ...627..286T} {627, 286}

\bibitem[\protect\citeauthoryear{{Tanaka}, {Takeuchi}  \& {Ward}}{{Tanaka}
  et~al.}{2002}]{Tanaka/Takeuchi/Ward/2002}
{Tanaka} H.,  {Takeuchi} T.,   {Ward} W.~R.,  2002, \apj, 565, 1257

\bibitem[\protect\citeauthoryear{{Testi} et~al.,}{{Testi}
  et~al.}{2014}]{Testi/etal/2014}
{Testi} L.,  et~al., 2014, Protostars and Planets VI, pp 339--361

\bibitem[\protect\citeauthoryear{{Throop} \& {Bally}}{{Throop} \&
  {Bally}}{2005}]{Throop/Bally/2005}
{Throop} H.~B.,  {Bally} J.,  2005, \apjl, 623, L149

\bibitem[\protect\citeauthoryear{{Ward}}{{Ward}}{1986}]{Ward/1986}
{Ward} W.~R.,  1986, \icarus, 67, 164

\bibitem[\protect\citeauthoryear{{Ward}}{{Ward}}{1997}]{Ward/1997}
{Ward} W.~R.,  1997, \icarus, 126, 261

\bibitem[\protect\citeauthoryear{{Youdin} \& {Goodman}}{{Youdin} \&
  {Goodman}}{2005}]{Youdin/Goodman/2005}
{Youdin} A.~N.,  {Goodman} J.,  2005, \apj, 620, 459

\bibitem[\protect\citeauthoryear{{Youdin} \& {Johansen}}{{Youdin} \&
  {Johansen}}{2007}]{Youdin/Johansen/2007}
{Youdin} A.,  {Johansen} A.,  2007, \apj, 662, 613

\bibitem[\protect\citeauthoryear{{Youdin} \& {Lithwick}}{{Youdin} \&
  {Lithwick}}{2007}]{Youdin/Lithwick/2007}
{Youdin} A.~N.,  {Lithwick} Y.,  2007, \icarus, 192, 588

\bibitem[\protect\citeauthoryear{{van Boekel}, {Min}, {Waters}, {de Koter},
  {Dominik}, {van den Ancker}  \& {Bouwman}}{{van Boekel}
  et~al.}{2005}]{van-Boekel/etal/2005}
{van Boekel} R.,  {Min} M.,  {Waters} L.~B.~F.~M.,  {de Koter} A.,  {Dominik}
  C.,  {van den Ancker} M.~E.,   {Bouwman} J.,  2005, \aap, \href
  {http://adsabs.harvard.edu/abs/2005A%26A...437..189V} {437, 189}

\makeatother
\end{thebibliography}



\appendix

\section{Derivation of the jump conditions}
\label{sec:jump_cond_derivation}

In our model, the gas within the disc flows vertically until it reaches an inclined ionisation front that separates the neutral disc from the  ionised wind. Across the front, the perpendicular velocity components $\vgnperp$ and $\vgiperp$ are constrained by the Rankine-Hugoniot equations in \cref{eq:RH_mass_conserv,eq:RH_mom_conserv} while the parallel components remain unchanged (i.e. $\vgipar = \vgnpar = \vgpar$). The inclination angle $\iIF$, the sound speed in the disc ($\cs$) and wind ($\csi$), and the ionised gas velocity ($\vgi$) and density ($\rhogi$) at the base of the wind are all known parameters of the model. This leaves a total of five unknown variables that need to be determined: $\vgn$, $\vgnperp$, $\vgpar$, $\vgiperp$, and $\rhogn$. Together with the jump conditions, we close the system of equations by using trigonometric relations between the velocity components:
\begin{align}
	\vgiperp^2 & = \vgi^2 - \vgpar^2,
	\label{eq:trig1}
\\
	\vgnperp & = \vgpar \cot{\iIF},
	\label{eq:trig2}
\\
	\vgn & = \vgpar \csc{\iIF}.
	\label{eq:trig3}
\end{align}

Using \cref{eq:RH_mass_conserv} to replace $\rhogn$ in \cref{eq:RH_mom_conserv}, we obtain a quadratic equation for $\vgnperp$,
\begin{equation}
	\vgiperp \vgnperp^2 - \left( \csi^2 + \vgiperp^2\right) \vgnperp + \cs^2 \vgiperp = 0,
\end{equation}
that is satisfied by \cref{eq:vgn}. Substituting this relation for $\vgnperp$ back into \cref{eq:RH_mass_conserv} and solving for $\rhogn$ yields (after rationalising the denominator) the second equality in \cref{eq:rhogn}. Meanwhile, by setting \cref{eq:vgn} equal to \cref{eq:trig2} and isolating the radical we find
\begin{equation}
	\csi^2+\vgiperp^2 -2 \vgiperp \vgpar \cot{\iIF} = \sqrt{\left( \csi^2 + \vgiperp^2 \right)^2 - 4 \cs^2 \vgiperp^2}.
\end{equation}
Squaring both sides, canceling like terms, and rearranging those remaining gives
\begin{equation}
	\left( \csi^2 + \vgiperp^2 \right) \vgpar \cot{\iIF} = \vgiperp \left( \cs^2 + \vgpar^2 \cot^2{\iIF} \right).
\end{equation}
Squaring the equation again, we then use \cref{eq:trig1} to eliminate $\vgiperp$ and acquire an equation for $\vgpar$ in terms of known variables. Finally, by expanding bracketed terms, collecting powers of $\vgpar$, and simplifying the coefficients we arrive at \cref{eq:vgpar,eq:c6,eq:c4,eq:c2,eq:c0} from the main text. Once $\vgpar$ is obtained, the remaining quantities are easily computed from the constraint equations.


\bsp	
\label{lastpage}
\end{document}